\documentclass{aa}

\usepackage{graphicx}
\usepackage{txfonts}
\usepackage{hyperref}
\usepackage{lscape}

\usepackage{natbib}
\bibpunct{(}{)}{;}{a}{}{,} 

\begin{document}

	\title{Chemical enrichment in LINERs from MaNGA. I. Tracing the nuclear abundances of oxygen and nitrogen in LINERs with varied ionizing sources}
	\titlerunning{LINERs from MaNGA. I. Nuclear abundances of oxygen and nitrogen}

	\author{Borja Pérez-Díaz\inst{\ref{inst1}}
		\and
		Enrique Pérez-Montero\inst{\ref{inst1}} \and Igor A. Zinchenko\inst{\ref{inst2}, \ref{inst3}} \and José M. Vílchez\inst{\ref{inst1}}}

	\institute{Instituto de Astrofísica de Andalucía (IAA-CSIC), Glorieta de la Astronomía s/n, 18008 Granada, Spain\label{inst1} \\\email{bperez@iaa.es}\and Faculty of Physics, Ludwig-Maximilians-Universit\"{a}t, Scheinerstr. 1, 81679 Munich, Germany\label{inst2} \and Main Astronomical Observatory, National Academy of Sciences of Ukraine, 27 Akad. Zabolotnoho St 03680 Kyiv, Ukraine\label{inst3}}

	\date{Received MONTH DAY, YEAR; accepted MONTH DAY, YEAR}



\abstract
{The chemical enrichment in low-ionization nuclear emission-line regions (LINERs) is still an issue with spatial resolution spectroscopic data because we lack studies and because the nature of their ionizing source is uncertain, although they are the most abundant type of active galaxies in the nearby Universe.}
{Considering different scenarios for the ionizing source (hot old stellar populations, active galactic nuclei, or inefficient accretion disks), we analyze the implications of these assumptions to constrain the chemical content of the gas-phase interstellar medium.}
{We used a sample of 105 galaxies from the survey called  Mapping Nearby Galaxies at Apache Point Observatory (MaNGA), whose nuclear central spaxels show LINER-like emission. For each scenario we considered, we built a grid of photoionization models (4928 models for each considered ionizing source) that were later used in the open-source code \textsc{HII-CHI-Mistry}. This allowed us to estimate chemical abundance ratios such as 12+log(O/H) or log(N/O) and to constrain the ionization parameters that characterize the ionized interstellar medium in these galaxies.}
{The oxygen abundances in the nuclear region of LINER-like galaxies are spread over a wide range 8.08 $<$ 12+log(O/H) $<$ 8.89, with a median solar value (in agreement with previous studies) when models for active galactic nuclei are considered. Nevertheless, the nitrogen-to-oxygen ratio we derived is much less affected by the assumptions on the ionizing source and indicates suprasolar  values (log(N/O) = -0.69). By comparing the different scenarios, we show that if hot old stellar populations caused the ionization of the interstellar medium, a complex picture (e.g., outflows and/or inflows that scale with the galaxy chemical abundance) would be needed to explain the chemical enrichment history, whereas the assumption of active galactic nucleus activity is compatible with the standard scenario that is found in most galaxies.}
{}


\maketitle



\section{Introduction}
\label{sec1}
Among the different approaches for analyzing the evolution of galaxies, the metal content of the interstellar medium (ISM) is a very important constraint. While Big Bang nucleosynthesis allows us to explain the majority of hydrogen (H), helium (He), lithium (Li), and deuterium (D) in the ISM \citep{Cyburt_2016}, nearly all the remaining elements are produced by means of the different star formation processes that govern the evolution of galaxies \citep[e.g.][]{Kobayashi_2020, Duarte-Puertas_2022}, which are ultimately shaped by other processes such as inflows, outflows, or merger events that affect the hydrostatic equilibrium within galaxies \citep[e.g.][]{Perez-Diaz_2024, Sharda_2024}.

Since the pioneering works from \citet{Peimbert_1967, Peimbert_1969, Searle_1971, Searle_1972}, the chemical content of the gas-phase ISM is generally traced by the oxygen abundance since it is the most abundant element in mass (\ensuremath{\sim}55\%, \citealt{Peimbert_2007}), and its presence helps to cool the nebula through emission lines that are detected in the ultraviolet (UV), optical (Opt), and infrared (IR) regimes. The brightest of these lines arise from collisional excitation. The oxygen abundance, expressed relative to hydrogen [12+log(O/H)], is affected by flow dynamics that can alter one or both quantities. Hence, complementary information is required to constrain the chemical enrichment stage of the ISM. In a simplified picture of the complex nucleosynthesis of elements such as nitrogen (N) and carbon (C), they are produced by massive stars (primary), but an additional channel of production arises when oxygen is already present in the ISM from which intermediate-mass stars were born, allowing them to produce N and C via CNO cycles \citep[secondary, e.g.][]{Henry_2000}. Complexity arises from other important aspects in chemical evolution models such as the time delay of ISM pollution from intermediate-mass stars \citep[e.g.][]{Vincenzo_2016}, the role of fast-rotator stars \citep[e.g.][]{Grisoni_2021}, or variations in the star formation histories \citep[e.g.][]{Molla_2006}. Thus, by analyzing the log(N/O) and/or log(C/O) ratios, we gain additional information to constrain the chemical enrichment history in galaxies. 

The presence of these elements in the gas-phase ISM helps to cool it by emission lines, which are used to estimate the chemical content. Recipes for estimating chemical abundances in the gas-phase ISM have been proposed, exploited, and perfected over the past decades. All of them can be summarized into three main groups. The first group relies on using auroral emission lines, such as [\ion{O}{iii}]\ensuremath{\lambda }4363\ensuremath{\AA}, [\ion{N}{ii}]\ensuremath{\lambda }5755\ensuremath{\AA} and/or [\ion{S}{iii}]\ensuremath{\lambda }6312\ensuremath{\AA} as well as other collisionally excited lines (CEls) to constrain the physical properties of the gas-phase ionized ISM such as the electron temperature (T\ensuremath{_{e}}) and density (n\ensuremath{_{e}}). This is called the direct method or T\ensuremath{_{e}}-method \citep[e.g.][]{Perez-Montero_2017}. Some problems that arise from this method are i) the faint nature of these auroral emission lines, ii) the lack of emission lines from high-ionized species (e.g., [OIV], [SIV], and [NeV]) in the optical range imposes an assumption on the ionization correction factors (ICFs), and iii) the fluctuations of the physical properties required, such as the temperature, might induce systematic uncertainties in their determination \citep{Martinez-Delgado_2023}, although some ratios such as log(N/O) are not affected by these fluctuations. The second technique relies on the use of photoionization models such as \textsc{cloudy} \citep{Ferland_2017}, \textsc{mappings} \citep{Sutherland_2017} , or \textsc{suma} \citep{Contini_1983}, in which the physical and chemical conditions in the gas-phase ISM as well as the source of ionization are given and are then constrained either to directly match the observations with customized models \citep[e.g.][]{Perez-Montero_2010} or by Bayesian approaches such as \textsc{HII-CHI-Mistry} \citep[\textsc{HCm},][]{Perez-Montero_2014}, \textsc{NebulaBayes} \citep{Thomas_2018}, or \textsc{Homerun} \citep{Marconi_2024}. Finally, a different technique that is more widely used because it is easy to apply is the use of strong emission-line ratios or combinations that correlate with the chemical abundances, which are calibrated by using one of the other two previous techniques. A summary of these calibrators can be found in Tab. 1 from \citet{Maiolino_2019}.

These techniques have been used over decades \citep{McClure_1968, Lequeux_1979, Vilchez_1988, Thuan_1995, Perez-Montero_2005, Andrews_2013, Fernandez-Ontiveros_2021}, especially for analyzing the chemical composition of the gas-phase ISM in star formation dominated galaxies (SFGs) over different cosmic times from either slit spectroscopy or integral field spectroscopy. However, to properly constrain the galaxy chemical evolution, the studies cannot focus on SFGs alone, but also need to study galaxies hosting active galactic nuclei (AGNs), which have an important role not only in galaxy evolution \citep[e.g.][]{Page_2004, Nicola_2019} because AGN feedback affects the gas cooling, and consequently, the star formation \citep{Morganti_2017, Capelo_2023}, but also in cosmological structures \citep[e.g.][]{Gitti_2012, Eckert_2021, Nobels_2022}. However, studies of the chemical content of the gas-phase ISM within galaxies hosting AGNs are rather rare. The gas-phase ISM surrounding the supermassive black hole (SMBH) in AGNs is divided into two different regions: the broad-line region (BLR), which is located close \citep[r \ensuremath{\sim} 0.01 pc][]{Mandal_2021} to the SMBH and is characterized by high densities (\ensuremath{n_{e} > 10^{9}} cm\ensuremath{^{-3}}) and broad emission-line components (\ensuremath{> 1000} km/s) \citep{Peterson_2006}; and the narrow-line region, which is located at greater distances \citep[r \ensuremath{\gtrsim} 1 pc][]{Peterson_2013} and has lower densities (\ensuremath{n_{e} \sim 10^{3}} cm\ensuremath{^{-3}}), and narrower emission-line components (\ensuremath{\sim 500} km/s) \citep{Peterson_2006}. The BLR presents several problems for the estimation of its abundance composition both observationally due to the high covering factors \citep{Gaskell_2009} and theoretically due to the complex gas dynamics governing its motion \citep{Gaskell_2009} or even the high variability in the broad emission lines that trace the BLR \citep{Ilic_2017}. These problems disappear for the NLR because the physical properties are very similar to those of HII regions.

Nevertheless, the determination of chemical abundances in the NLR presents some caveats. It was shown by \citet{Dors_2015} that when the direct method is used to constrain the chemical abundances, unexpectedly low chemical abundances are obtained that are not found with other techniques. Moreover, the lack of a detection in the optical of emission lines from highly ionized species, which are expected because the ionizing front from the AGN is harder \citep{Perez-Diaz_2022, Perez-Diaz_2024}, adds more uncertainty to its estimates. Hence, most of the techniques for studying chemical abundances in the NLR of AGNs rely on either using photoionization models to directly estimate their abundances or on providing calibrations to estimate them in larger samples \citep[e.g.][]{Perez-Montero19, Carvalho_2020, Perez-Diaz_2021, Perez-Diaz_2022, Dors_2023}. 

An additional bias in the studies of chemical abundances in AGNs is that they mainly target Seyferts 2, that is, highly ionized AGNs, whereas low-luminosity AGNs (LLAGNs), such as LINERs, are the most common type of AGN in the local Universe \citep{Ho_1997}. Studies of chemical abundances in LLAGNs are even rarer. One of the main reasons for this rarity is the fact that there is no consensus in the literature about the source of ionization of LINERs. Possible sources include i) sub-Eddington accretion \citep{Kewley_2006, Ho_2009} onto SMBHs, which can later affect the flow of gas by even truncating the disk through advected-dominated flows \citep[ADAFs,][]{Nemmen_2014}; ii) hot old stellar populations that are dominated by post-asymptotic giant branch (pAGB) stars \citep{Binette_1994, Stasinska_2008}; and iii) fast radiative shocks \citep{Dopita_1995, Allen_2008}.

\citet{Annibali_2010} targeted 67 early-type galaxies in the local Universe in one of the first works that estimated the oxygen abundances in a statistically significant sample of LLAGNs, since of these galaxies (\ensuremath{\sim 72}\%) hosted LINERs. However, the authors relied on calibrations that were obtained from photoionization models assuming AGNs \citep{Storchi_1998} as the source of ionization, or on empirical calibrations taken from the analysis of early-type galaxies, which would mimic the conditions in their sample \citep{Kobulnicky_1999}. Their resulting oxygen abundances were in the range 8.49 $<$ 12+log(O/H) $<$ 9.01, and most of them were slightly suprasolar\footnote{Hereinafter, we assume the solar abundances reported by \citet{Asplund_2009}, that is, 12+log(O/H)\ensuremath{_{\odot}}=8.69, 12+log(N/H)\ensuremath{_{\odot}}=7.83, and log(N/O)\ensuremath{_{\odot}}=-0.86}.  

More recently, \citep{Perez-Diaz_2021} systematically analyzed 16 Seyferts 2 and 40 LINERs from the Palomar Spectroscopic Survey \citep{Ho_1997}, complemented by an additional sample of 25 LINERs from \citep{Povic_2016}. To estimate the oxygen abundances, the authors employed \textsc{HII-CHI-Mistry} \citep[hereinafter \textsc{HCm},][]{Perez-Montero_2014, Perez-Montero19}, a Bayesian-like code that relies on a large grid of photoionization models, assuming AGNs as the source of ionization, because they found that multiwavelength studies indicated AGN feature emissions in X-rays and radio wavelengths. The authors reported that some (\ensuremath{\sim} 15\%) LINERs were characterized by unexpectedly low oxygen abundances (12+log(O/H) $<$ 8.3). Later on, \citet{Oliveria_2022} presented an analysis of the nuclear region of 43 LINERs from the SDSS IV - MaNGA survey \citep{Bundy_2015, Blanton_2017} from photoionization models assuming pAGB stars as the source of ionization. They concluded that their sample of LINERs was characterized by solar-like oxygen abundances of 8.54 $<$ 12+log(O/H) $<$ 8.84. A similar result was found in a recent study by \citet{Oliveira_2024b}, who reported that weak AGNs present oxygen abundances in the range 8.50 $<$ 12+log(O/H) $<$ 8.90. Only the study by \citep{Krabbe_2021}, who only analyzed one LINER galaxy (UGC 4805) based on MaNGA data and photoionization models, simultaneously explored the AGN and the pAGBs ionizing scenarios. While their finding of metallicity trends did not provide a conclusive clue, they suggested that the pAGB scenario is the most favorable based on the position of this nucleus on some specific  diagnostic diagrams for the observed emission lines, as well as on the observed high degree of ionization. 

Nitrogen abundances in LLAGNs have been analyzed even more infrequently. Except for the pioneering work of \citet{Perez-Diaz_2021}, only the recent work by \citet{Oliveira_2024} analyzed the log(N/O) abundance ratio and its relation with the 12+log(O/H) abundance in order to constrain the chemical enrichment of these objects. Although both studies employed photoionization models and assumed different ionizing sources (AGNs and pAGBs, respectively), the results were consistent. This shows that LINERs tend to have slightly higher log(N/O) than is reported in SFGs.

Exploiting the capabilities of integral field spectroscopic (IFS) data from the SDSS IV - MaNGA survey \citep{Bundy_2015, Blanton_2017}, we present a series of papers that analyze the spatially resolved chemical enrichment of the gas-phase ISM in galaxies hosting LINER-like nuclear emission. We consider for the first time different scenarios that might explain the observed emission in a statistically significant sample of galaxies. This first paper of the series focuses on the determination of nuclear chemical abundances in this sample, and it is organized as follows. In section \ref{sec2} we explain the selection of the sample of galaxies hosting LINER-like emission. We also introduce other different criteria we used in this study. In section \ref{sec3} we discuss the method we employed to estimate the chemical abundances in our sample. In section \ref{sec4} we present the main results of this study, and we discuss them in section \ref{sec5}. In section \ref{sec6} we summarize the main conclusions. We assumed the cosmological parameters given by \ensuremath{\Omega_{m} = 0.3}, \ensuremath{\Omega_{\Lambda} = 0.7}, and \ensuremath{H_{0} = 67} km/s/Mpc.

\section{Sample selection}
\label{sec2}
\subsection{MaNGA data and emission-line measurements} 
The Mapping Nearby Galaxies at Apache Point Observatory \citep[MaNGA;][]{Bundy_2015} is part of the Sloan Digital Sky Survey IV \citep[SDSS IV;][]{Blanton_2017}. For this work, we used data release 17 (DR17, \citealt{Abdu_2022}). We used individual spaxels whose size was significantly smaller than that of the point spread function (PSF) in the MaNGA datacubes. The spatial resolution of these cubes has a median full width at half maximum (FWHM) of 2.54 arcsec \citep{Law_2016}. 

We examined the MaNGA spectra as outlined in \citet{Zinchenko2016,Zinchenko2021}. Briefly, we used the code STARLIGHT \citep{CidFernandes2005,Mateus2006,Asari2007} to fit the stellar background across all spaxels, adapting it for parallel datacube processing. Simple stellar population (SSP) spectra from \citet{Bruzual_2003} evolutionary synthesis models were used for stellar fitting, and we subtracted them from the observed spectrum for a pure gas spectrum. Then, we fit emission lines using our ELF3D code. Each emission line was fit with a single-Gaussian profile. For each spectrum, we measured the fluxes of the [\ion{O}{II}]\ensuremath{\lambda,\lambda}3726,3729\ensuremath{\AA} (hereinafter [\ion{O}{II}]\ensuremath{\lambda}3727\ensuremath{\AA}), [\ion{Ne}{III}]\ensuremath{\lambda}3868\ensuremath{\AA}, H\ensuremath{_{\beta}}, [\ion{O}{III}]\ensuremath{\lambda}4959\ensuremath{\AA}, [\ion{O}{III}]\ensuremath{\lambda}5007\ensuremath{\AA}, [\ion{N}{II}]\ensuremath{\lambda}6548\ensuremath{\AA}, H\ensuremath{_{\alpha}}, [\ion{N}{II}]\ensuremath{\lambda}6584\ensuremath{\AA}, and [\ion{S}{II}]\ensuremath{\lambda,\lambda}6717,6731\ensuremath{\AA} lines with a signal-to-noise ratio, S/N, \ensuremath{>} 3.

\subsection{Sample classification}
From the original sample of galaxies presented in the SDSS IV - MaNGA survey, we initially selected the galaxies that were classified as LINERs in at least one of the three diagnostic (BPT) diagrams \citep{Baldwin_1981, Veilleux_1987} with the semi-empirical constraints proposed by \citet{Kauffmann_2003} and \citet{Kewley_2006}. For this purpose, we used the emission lines H\ensuremath{_{\beta}}, [\ion{O}{iii}]\ensuremath{\lambda }5007\ensuremath{\AA}, [\ion{O}{i}]\ensuremath{\lambda }6300\ensuremath{\AA}, [\ion{N}{ii}]\ensuremath{\lambda }6584\ensuremath{\AA}, [\ion{S}{ii}]\ensuremath{\lambda }6717\ensuremath{\AA}, and [\ion{S}{ii}]\ensuremath{\lambda }6731\ensuremath{\AA}. Additionally, we complemented the diagrams with information on the equivalent width (EW) for H\ensuremath{_{\alpha}}, as proposed by \citet{Cid-Fernandes_2011}, since the strength of these feature allowed us to better distinguish between galaxies whose activity is dominated by strong AGNs (sAGN; \ensuremath{W_{H_{\alpha}} > 6\AA}), weak AGNs (wAGN; \ensuremath{3\AA $<$ W_{H_{\alpha}} $<$ 6\AA}), and retired galaxies (RG; \ensuremath{0.5\AA $<$ W_{H_{\alpha}} $<$ 3\AA}). Below this limit lies the region that is dominated by passive galaxies (PG; \ensuremath{W_{H_{\alpha}} $<$ 3\AA}), which are mainly line-less galaxies \citep{Cid-Fernandes_2011}.

To perform the classification, we focused our attention on the central spaxel for each galaxy. The median distance for our sample of galaxies is 150.56 Mpc within the range of [35.24 Mpc, 560.09 Mpc]. This implies that the angular sampling is 1.460 kpc on average within the range of [0.342 kpc, 5.4 kpc] for the 2\ensuremath{^{\prime\prime}} fiber coverage in the sky \citep{Bundy_2015}. This allowed us to capture most of the emission when an AGN is the source of ionization, considering the average sizes of the narrow- line region \citep[\ensuremath{\sim } 1-5 kpc][]{Bennert_2006a, Bennert_2006b}. This range can also accommodate the case in which pAGB stars ionize the gas-phase ISM \citep[e.g.][]{Binette_1994}.  While other authors integrated the flux in a circular aperture with a radius of 1 kpc \citep{Oliveira_2024}, we obtained that the average change in the emission-line flux ratios is smaller than 5\%, and the classification and later results are therefore not affected by selecting either the central spaxel or an aperture of 1 kpc.

\begin{figure*}
	\centering
	\includegraphics[width=0.9\hsize]{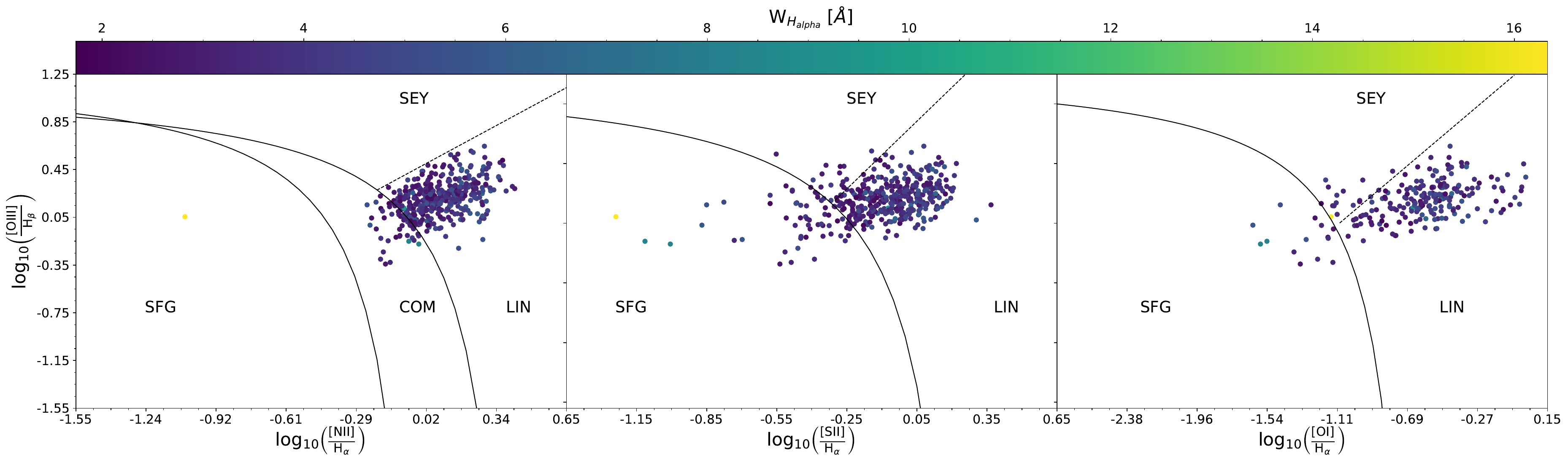}
	\caption{Diagnostic diagrams of the central spaxels in our sample of galaxies. The color bar shows the equivalent width for H\ensuremath{_{\alpha}} (\ensuremath{W_{H_{\alpha}}}). The solid and dashed lines represent the region limits as defined by \citet{Kewley_2006}, with the exception of the separation between Seyfert 2 and LINERs in the [NII]/H\ensuremath{_{\alpha}} diagram, which was taken from \citet{Cid-Fernandes_2010}. Each region is labeled as follows: SFG for star-forming galaxies, SEY for Seyferts, LIN for LINERs, and COM for composites.}
	\label{diagnostic}
\end{figure*}

\begin{figure}
	\centering
	\includegraphics[width=0.9\hsize]{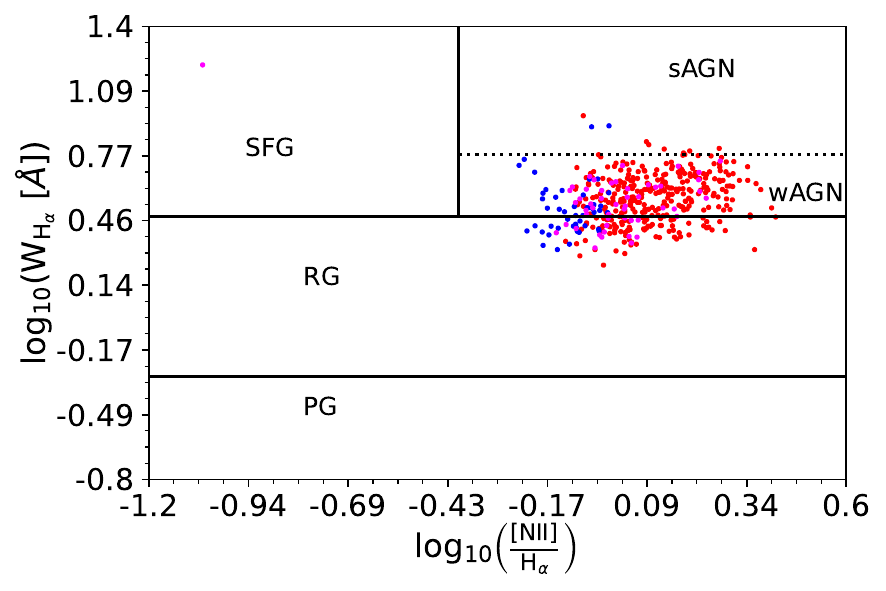}
	\caption{WHAN diagram, showing the region demarcations as provided by \citet{Cid-Fernandes_2010, Cid-Fernandes_2011}. The color code represents the classification as given by the BPT diagrams (see Figure \ref{diagnostic}): Red dots represent galaxies classified as LINERs, blue dots represent galaxies falling in the composite region, and magenta dots show galaxies without a clear classification. Each region is labeled as follows: SFG for star-forming galaxies, sAGN for strong AGNs, wAGN for weak AGNs, RG for retired galaxies, and PG for passive galaxies.}
	\label{whan_diagnostic}
\end{figure}

We show in Figure \ref{diagnostic} the diagnostic diagrams for the central spaxels in our sample of galaxies from MaNGA. When we only consider the galaxies that are classified by at least one of the diagrams as LINERs, the sample contains 429 galaxies. Considering that problems of contamination from diffuse ionized gas (DIG) in [\ion{S}{ii}] are higher than in [\ion{N}{II}] lines \citep{Perez-Montero_2023}, we imposed the condition that galaxies are simultaneously classified as LINERs in all three diagrams. This reduced the sample to 329 galaxies, with values of \ensuremath{W_{H_{\alpha}}} in the range [1.74\ensuremath{\AA}, 9.25\ensuremath{\AA}]. We cannot rule out the possibility of contamination from different ionizing sources \citep[e.g. the combination of star formation and AGN activity,][]{Davies_2016}, and we therefore complemented our classification with information from the WHAN diagram \citep{Cid-Fernandes_2011}. As shown in Figure \ref{whan_diagnostic}, the majority of our galaxies fall in the wAGN region, and some galaxies present slightly higher or lower values. Only one galaxy is classified as SFG due to the weak ratio of \ensuremath{H_{\alpha}} and [\ion{N}{ii}]\ensuremath{\lambda }6584\ensuremath{\AA}. Table \ref{distributions} shows that the majority of the sample falls in the wAGN region (300; \ensuremath{\sim 70}\%). When we focus our attention only on the galaxies that are classified as LINERs based on the BPT diagram (329; \ensuremath{\sim 77}\%), the percentage of wAGN-like galaxies is even higher than in the overall sample (\ensuremath{\sim 73}\%).

\begin{table*}
\caption{Classification of the central spaxels from the preliminary sample of 429 galaxies selected from MaNGA.}
\label{distributions}
\centering
\begin{tabular}{c | c c c c c c} 
\hline\hline
Classification & SFG & sAGN & wAGN & RG & PG & Total \\ \hline 
LINERs & 0 (0.00) & 7 (2.13) & 239 (72.64) & 83 (25.23) & 0 (0.00) & 329 (76.87) \\ 
Composites & 0 (0.00) & 2 (4.17) & 26 (54.17) & 20 (41.67) & 0 (0.00) & 48 (11.21) \\ 
Ambiguous & 1 (1.96) & 0 (0.00) & 35 (68.63) & 15 (29.41) & 0 (0.00) & 51 (11.92) \\ \hline
Total & 1 (0.23) & 9 (2.10) & 300 (70.09) & 118 (27.57) & 0 (0.00) & 428 (100.00) \\ 
\end{tabular}
\tablefoot{The headline row shows the classifications based on the WHAN diagram \citep{Cid-Fernandes_2010, Cid-Fernandes_2011}. The first column shows the classification based on the BPT diagrams \citep{Kauffmann_2003, Kewley_2006}. In parentheses, we provide the relative percentage per group.}
\end{table*}

Another constraint we added to the sample selection is that the detected LINER-like emission mainly comes from the central spaxel. By restricting our sample with this criterion, we ensured i) that there are enough HII regions to analyze spatially resolved properties in all galaxies, such as metallicity gradients (second paper of these series), ii) that the origin of the LINER-like emission cannot be automatically excluded to be an AGN, and iii) that we can test for the same galaxy what occurs in their SF-dominated regions as opposed to the nuclear region. By adding this constraint, we obtained a sample of 105 galaxies, all of which show LINER-like emission according to all BPT diagrams for the central spaxels. Of these, 57 are classified as wAGN and 48 as retired galaxies according to the WHAN diagram \citep{Cid-Fernandes_2010, Cid-Fernandes_2011}.
\section{Determining the chemical abundance}
\label{sec3}
In this section, we present a detailed explanation of the method we used to estimate the chemical abundances as well as other physical parameters from the nebular emission that we retrieved from the MaNGA data. Because we accounted for possible different natures of the ionizing source in the nuclear regions of our sample of galaxies, we simultaneously  analyzed all these scenarios by means of photoionization models.
\subsection{HII-CHI-Mistry}
To estimate the chemical abundances, we relied on the Bayesian-like Python code \textsc{HII-CHI-Mistry}\footnote{The code is publicly available at \url{http://home.iaa.csic.es/~epm/HII-CHI-mistry.html}.} (or \textsc{HCm}), which was originally developed by \citet{Perez-Montero_2014} for HII regions, but was later extended to AGN sources \citep{Perez-Montero19, Perez-Diaz_2021} and extreme emission-line galaxies \citep{Perez-Montero_2021} for the optical version. We made use of the optical version 5.5 of the code. \textsc{HCm} uses a grid of photoionization models with three free parameters: the chemical properties of the gas-phase ISM 12+log(O/H) and log(N/O), and the ionization parameter log(U), estimated by comparing emission line ratios that are sensitive to these parameters. Other important properties for the computation of photoionization models such as the geometry, the density, the dust-to-gas ratio, and the stopping criteria were explored for each grid of models. In a first iteration, the code estimates log(N/O) and uses this estimate to constrain the grid of models. This first estimation was always done because we have all necessary emission lines for this determination (e.g., [\ion{O}{ii}]\ensuremath{\lambda }3727\ensuremath{\AA}, and [\ion{N}{ii}]\ensuremath{\lambda }6584\ensuremath{\AA}). In a second iteration, the code performs an estimation of 12+log(O/H) and log(U). Since all these parameters are free in the grid of models, all quantities were estimated independently of each other, and no relation was assumed between them. 

As input for \textsc{HCm}, we used the emission-line ratios  [\ion{O}{ii}]\ensuremath{\lambda }3727\ensuremath{\AA}, [\ion{Ne}{iii}]\ensuremath{\lambda }3868\ensuremath{\AA}, [\ion{O}{iii}]\ensuremath{\lambda }5007\ensuremath{\AA}, [\ion{N}{ii}]\ensuremath{\lambda }6584\ensuremath{\AA}, and [\ion{S}{ii}]\ensuremath{\lambda }6717\ensuremath{\AA}+[\ion{S}{ii}]\ensuremath{\lambda }6731\ensuremath{\AA}, referred to H\ensuremath{_{\beta}} emission. All emission-line ratios were corrected for reddening assuming case B photoionization and an expected ratio of H\ensuremath{_{\alpha}} and H\ensuremath{_{\beta}} of 3.1 for standard conditions in the NLR, that is, an electron density n\ensuremath{_{e} \sim 500} cm\ensuremath{^{-3}} and an electron temperature T\ensuremath{_{e} \sim 10^{4}} K \citep{Osterbrock_book}, and the extinction curve from \citet{Howarth_1983} for R\ensuremath{_{V}} = 3.1.

\subsection{Grids of photoionization models}
Whereas \textsc{HCm} allowed us to perform the same method regardless of the source of ionization, the user must select the grid of photoionization models to be used in the estimation. The differences therefore emerge based on the assumed spectral energy distribution (SED). All models were computed using \textsc{Cloudy}\footnote{The code is publicly available at \url{https://gitlab.nublado.org/cloudy/cloudy}.} v17 \citep{Ferland_2017}

In the case of HII regions, we chose a star formation cluster with an age of 1 Myr as the ionization source. We used the \textsc{POPSTAR} \citep{Molla_2009} synthesis code for an initial mass function (IMF) that followed the trend reported by \citet{Chabrier_2003}. The gas density was assumed to be 100 cm\ensuremath{^{-3}}. By allowing variation in the ionization parameter and metallicity, \citet{Perez-Montero_2014} demonstrated that these models are able to reproduce the emission-line ratios observed in HII regions.

For the nuclear emission, we built different grids of photoionization models that accounted for all proposed scenarios to reproduce the LINER-like emission. We list them below.
\begin{itemize}
\item Active galactic nucleus SEDs composed of two components: The first component represents the big blue bump, which peaks at 1 Ryd, and the other component represents a power law with a spectral index \ensuremath{\alpha_{X} = -1} that represents the nonthermal X-rays radiation. To test different regimes in the hardness of the ionization, we selected several values for the slope in the power law that traced the continuum between 2.5 keV and 2500 \ensuremath{\AA} (\ensuremath{\alpha_{OX}}) given by \ensuremath{\alpha_{OX} = } [-0.8, -1.0, -1.2, -1.4, -1.6, -1.8, -2.0]. Although it was reported that slopes below \ensuremath{\alpha_{OX} = -1.4} do not reproduce the observed emission-line ratios well \citep{Carvalho_2020, Oliveira_2024b}, we explored all these scenarios to search for significant differences caused by the choice of this parameter.
\item Post-AGB SEDs obtained from the grid of nonlocal thermodynamic equilibrium (NLTE) model atmospheres\footnote{They can be downloaded from \url{http://astro.uni-tuebingen.de/~rauch/TMAF/flux_H-Ca.html}.} provided by \citep{Rauch_2003} for three different effective temperatures: 5\ensuremath{\cdot 10^{4}} K, 1\ensuremath{\cdot 10^{5}} K and 1.5\ensuremath{\cdot 10^{5}} K. These SEDs were computed assuming \ensuremath{\log g = 6}, and we assumed normalized fractions of 0.33 for helium, 0.50 for carbon, 0.02 for nitrogen, and 0.15 for oxygen.
\item An advection-dominated accretion flow (ADAF) model for the AGN, which represents the truncation of the accretion disk due to inefficiency in the accretion process. In particular, we made use of the code \textsc{riaf-sed}\footnote{The code is publicly available at \url{https://github.com/rsnemmen/riaf-sed}.} \citep{Yuan_2005, Yuan_2007, Nemmen_2014}. Considering the large number of free parameters for the computation, we used the average SED that characterized ADAF emission in LLAGN \citep[see][for more details]{Nemmen_2014}.
\end{itemize}

For each considered SED, we modeled the ISM by allowing the oxygen abundance to vary in the range 6.9 $<$ 12+log(O/H) $<$ 9.1 in steps of 0.1 dex, the nitrogen-to-oxygen ratio varied in the range -2.0 $<$ log(N/O) $<$ 0.0 in steps of 0.125 dex, and the ionization parameter varied in the range -4.0 $<$ log(U) $<$ -0.5 in steps of 0.25 dex. This last parameter was preliminarily constrained in the range -4.0 $<$ log(U) $<$ -2.5 based on previous findings \citep[e.g.][]{Perez-Diaz_2021, Oliveria_2022, Oliveira_2024} because it was needed to break the degeneracy reported in some emission line ratios \cite{Perez-Montero19, Perez-Diaz_2021}. For all models, a density of 500 cm\ensuremath{^{-3}} was assumed. In total, we computed 4928 models per ionizing source, which yields a total of 54208 models. We also explored the effects of the stopping criteria by considering two different scenarios: for a ratio of ionized hydrogen atoms of 0.98 and of 0.02. As already reported in other studies \citep[e.g.][]{Perez-Diaz_2021, Perez-Diaz_2022}, these two scenarios do not introduce significant differences (they are lower (\ensuremath{\sim } 0.05 dex) than the steps in the grid of models) in the chemical abundance estimations.

\begin{table*}
\caption{Properties of the resulting distribution of the derived oxygen abundances.}
\label{distributions_OH}
\centering
\begin{tabular}{c | c c c c c } 
\hline\hline
\textbf{Model} & \textbf{Class.} & \textbf{12+log(O/H)}$_{med}$ & \textbf{12+log(O/H)}$_{sd}$ & \textbf{12+log(O/H)}$_{min}$ & \textbf{12+log(O/H)}$_{max}$  \\ \textbf{(1)} & \textbf{(2)} & \textbf{(3)} & \textbf{(4)} & \textbf{(5)} & \textbf{(6)} \\  \hline 
 & wAGN &  8.72 & 0.05 & 8.59 & 8.84 \\ 
AGN $\alpha_{OX}=-0.8$ & RG & 8.71 & 0.08 & 8.36 & 8.82 \\ 
 & All & 8.72 & 0.06 & 8.36 & 8.84 \\ \hline 
 & wAGN &  8.66 & 0.04 & 8.54 & 8.77 \\ 
AGN $\alpha_{OX}=-1.0$ & RG & 8.64 & 0.07 & 8.36 & 8.76 \\ 
 & All & 8.65 & 0.06 & 8.36 & 8.77 \\ \hline 
 & wAGN &  8.57 & 0.05 & 8.47 & 8.68 \\ 
AGN $\alpha_{OX}=-1.2$ & RG & 8.59 & 0.04 & 8.46 & 8.64 \\ 
 & All & 8.58 & 0.04 & 8.46 & 8.68 \\ \hline 
 & wAGN &  8.70 & 0.05 & 8.57 & 8.79 \\ 
AGN $\alpha_{OX}=-1.4$ & RG & 8.66 & 0.07 & 8.37 & 8.75 \\ 
 & All & 8.68 & 0.07 & 8.37 & 8.79 \\ \hline 
 & wAGN &  8.70 & 0.05 & 8.59 & 8.78 \\ 
AGN $\alpha_{OX}=-1.6$ & RG & 8.64 & 0.08 & 8.32 & 8.78 \\ 
 & All & 8.69 & 0.07 & 8.32 & 8.78 \\ \hline 
 & wAGN &  8.70 & 0.05 & 8.55 & 8.78 \\ 
AGN $\alpha_{OX}=-1.8$ & RG & 8.64 & 0.08 & 8.31 & 8.76 \\ 
 & All & 8.68 & 0.07 & 8.31 & 8.78 \\ \hline 
 & wAGN &  8.70 & 0.05 & 8.54 & 8.79 \\ 
AGN $\alpha_{OX}=-2.0$ & RG & 8.64 & 0.08 & 8.30 & 8.76 \\ 
 & All & 8.69 & 0.08 & 8.30 & 8.79 \\ \hline 
 & wAGN &  8.47 & 0.02 & 8.41 & 8.55 \\ 
pAGB $T_{eff}= 5\cdot10^{4}$ K & RG & 8.47 & 0.02 & 8.44 & 8.53 \\ 
 & All & 8.47 & 0.02 & 8.41 & 8.55 \\ \hline 
 & wAGN &  8.64 & 0.11 & 8.28 & 8.74 \\ 
pAGB $T_{eff}= 1\cdot10^{5}$ K & RG & 8.50 & 0.14 & 8.07 & 8.74 \\ 
 & All & 8.57 & 0.13 & 8.07 & 8.74 \\ \hline 
 & wAGN &  8.56 & 0.16 & 8.25 & 8.82 \\ 
pAGB $T_{eff}= 1.5\cdot10^{5}$ K & RG & 8.41 & 0.16 & 8.08 & 8.72 \\ 
 & All & 8.50 & 0.17 & 8.08 & 8.82 \\ \hline 
 & wAGN &  8.51 & 0.04 & 8.43 & 8.60 \\ 
ADAF & RG & 8.50 & 0.04 & 8.42 & 8.62 \\ 
 & All & 8.51 & 0.04 & 8.42 & 8.62 \\ \hline 
\end{tabular}
\tablefoot{For each photoionization model (1) and each group of LINERs (2) the median value (3), standard deviation (4), minimum (5) and maximum values (6) are provided.}
\end{table*}

\begin{figure*}
	\centering
	\includegraphics[width=0.9\hsize]{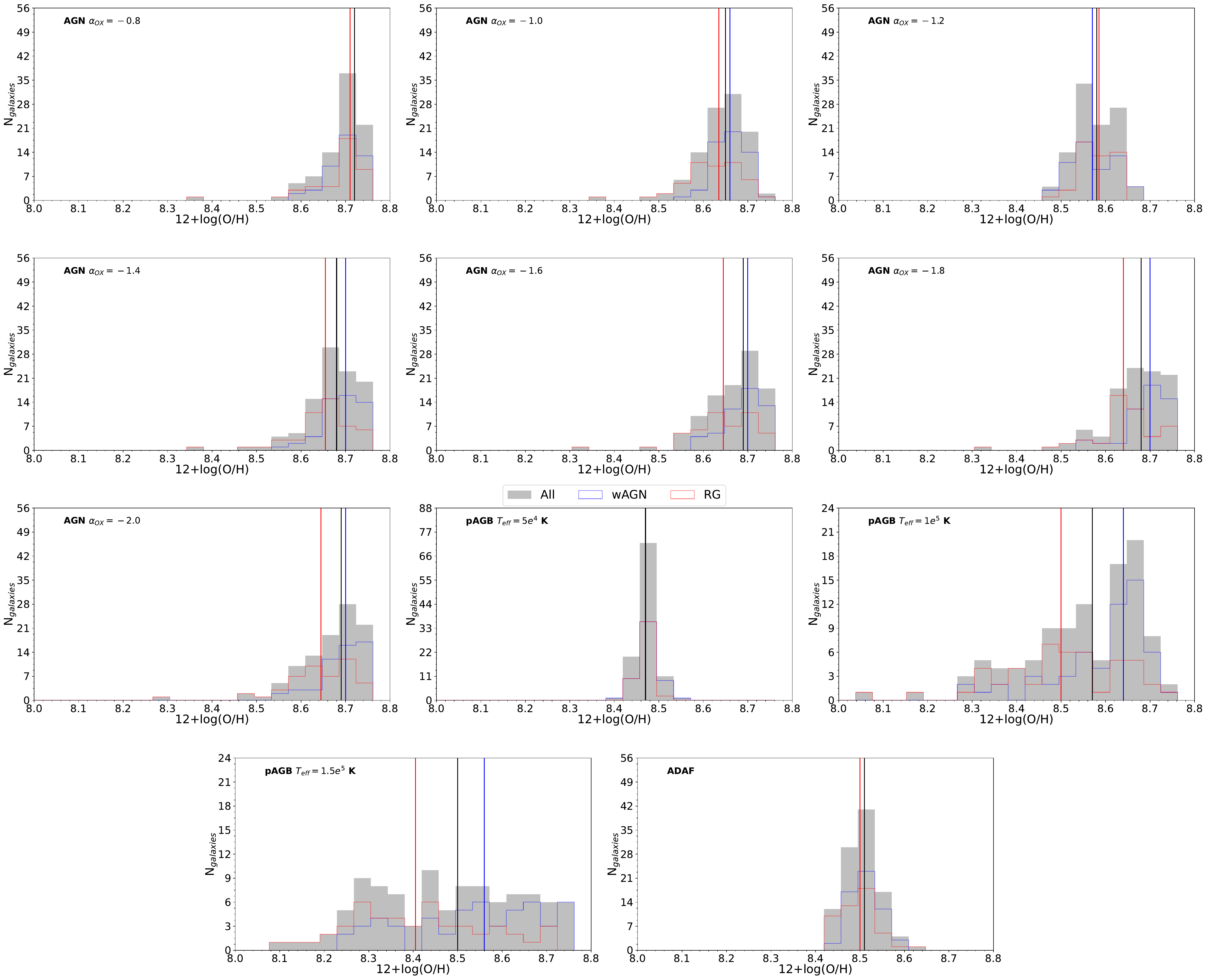}
	\caption{Histograms for 12+log(O/H) nuclear abundance ratios for LINERs for each photoionization grid of models. The step-filled gray histogram corresponds to all LINERs in our sample. The step blue and red histograms correspond to wAGNs and RGs, respectively. The vertical solid lines represent the median values for each distribution.}
	\label{OH_histogram}
\end{figure*}

\begin{table*}
\caption{Same as Table \ref{distributions_OH}, but for the nitrogen-to-oxygen abundance ratios log(N/O).}
\label{distributions_NO}
\centering
\begin{tabular}{c | c c c c c } 
\hline\hline
\textbf{Model} & \textbf{Class.} & \textbf{log(N/O)}$_{med}$ & \textbf{log(N/O)}$_{sd}$ & \textbf{log(N/O)}$_{min}$ & \textbf{log(N/O)}$_{max}$  \\ \textbf{(1)} & \textbf{(2)} & \textbf{(3)} & \textbf{(4)} & \textbf{(5)} & \textbf{(6)} \\  \hline 
 & wAGN &  -0.88 & 0.16 & -1.20 & -0.39 \\ 
AGN $\alpha_{OX}=-0.8$ & RG & -0.79 & 0.14 & -1.15 & -0.47 \\ 
 & All & -0.84 & 0.15 & -1.20 & -0.39 \\ \hline 
 & wAGN &  -0.81 & 0.15 & -1.14 & -0.42 \\ 
AGN $\alpha_{OX}=-1.0$ & RG & -0.74 & 0.14 & -1.12 & -0.43 \\ 
 & All & -0.78 & 0.15 & -1.14 & -0.42 \\ \hline 
 & wAGN &  -0.82 & 0.15 & -1.17 & -0.39 \\ 
AGN $\alpha_{OX}=-1.2$ & RG & -0.73 & 0.14 & -1.08 & -0.44 \\ 
 & All & -0.79 & 0.15 & -1.17 & -0.39 \\ \hline 
 & wAGN &  -0.83 & 0.15 & -1.19 & -0.40 \\ 
AGN $\alpha_{OX}=-1.4$ & RG & -0.75 & 0.14 & -1.11 & -0.50 \\ 
 & All & -0.79 & 0.15 & -1.19 & -0.40 \\ \hline 
 & wAGN &  -0.84 & 0.16 & -1.22 & -0.38 \\ 
AGN $\alpha_{OX}=-1.6$ & RG & -0.74 & 0.13 & -1.11 & -0.48 \\ 
 & All & -0.79 & 0.15 & -1.22 & -0.38 \\ \hline 
 & wAGN &  -0.83 & 0.15 & -1.18 & -0.38 \\ 
AGN $\alpha_{OX}=-1.8$ & RG & -0.75 & 0.14 & -1.12 & -0.47 \\ 
 & All & -0.81 & 0.15 & -1.18 & -0.38 \\ \hline 
 & wAGN &  -0.83 & 0.15 & -1.19 & -0.38 \\ 
AGN $\alpha_{OX}=-2.0$ & RG & -0.75 & 0.14 & -1.11 & -0.47 \\ 
 & All & -0.80 & 0.15 & -1.19 & -0.38 \\ \hline 
 & wAGN &  -0.63 & 0.13 & -0.93 & -0.33 \\ 
pAGB $T_{eff}= 5\cdot10^{4}$ K & RG & -0.56 & 0.12 & -0.88 & -0.33 \\ 
 & All & -0.58 & 0.13 & -0.93 & -0.33 \\ \hline 
 & wAGN &  -0.71 & 0.15 & -1.05 & -0.24 \\ 
pAGB $T_{eff}= 1\cdot10^{5}$ K & RG & -0.62 & 0.13 & -1.00 & -0.34 \\ 
 & All & -0.67 & 0.15 & -1.05 & -0.24 \\ \hline 
 & wAGN &  -0.71 & 0.15 & -1.06 & -0.27 \\ 
pAGB $T_{eff}= 1.5\cdot10^{5}$ K & RG & -0.64 & 0.14 & -1.00 & -0.37 \\ 
 & All & -0.70 & 0.15 & -1.06 & -0.27 \\ \hline 
 & wAGN &  -0.74 & 0.15 & -1.07 & -0.27 \\ 
ADAF & RG & -0.65 & 0.13 & -0.99 & -0.36 \\ 
 & All & -0.70 & 0.15 & -1.07 & -0.27 \\ \hline 
\end{tabular}
\end{table*}

\begin{figure*}
	\centering
	\includegraphics[width=0.9\hsize]{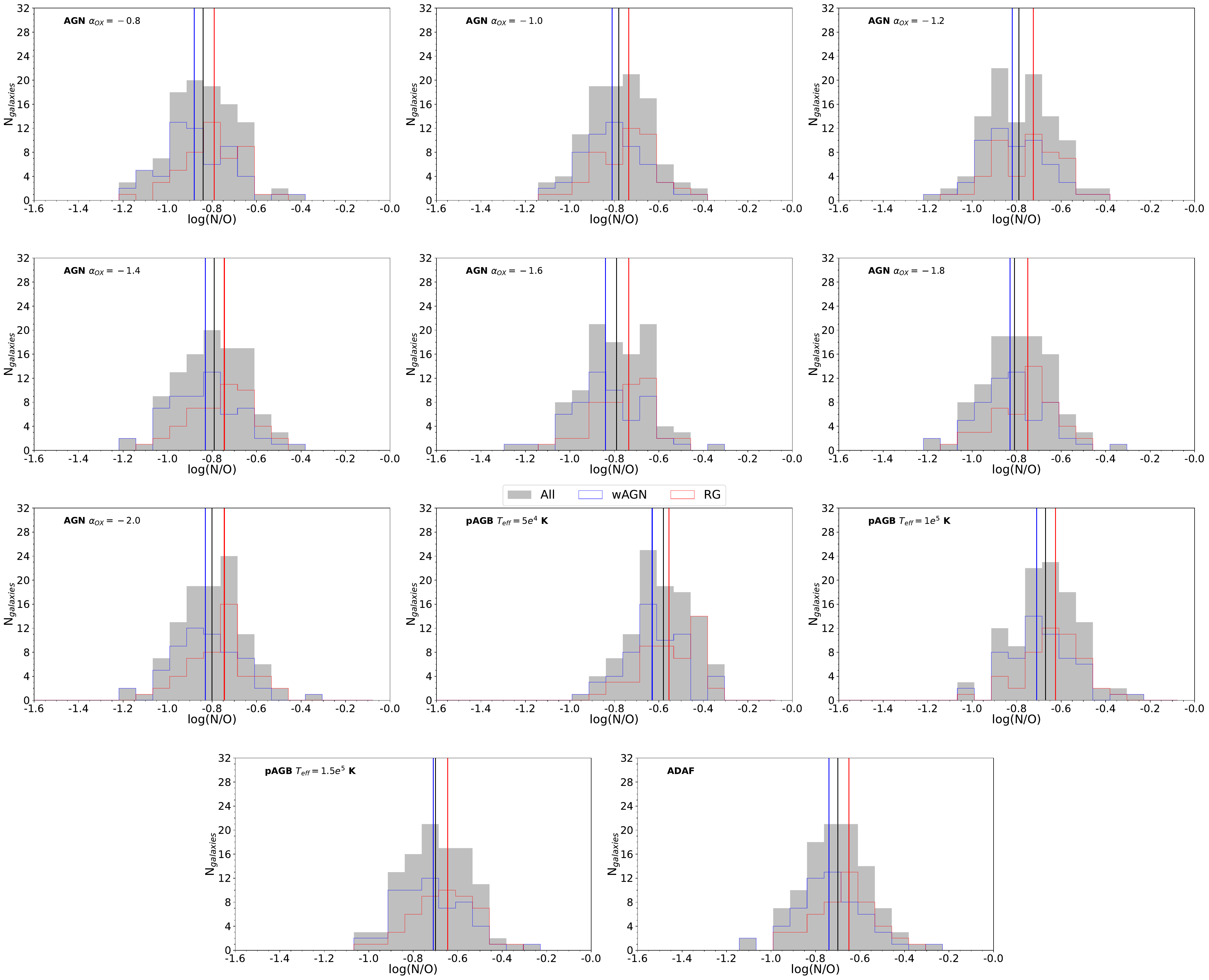}
	\caption{Same as Fig. \ref{OH_histogram}, but for log(N/O) nuclear abundance ratios.}
	\label{NO_histogram}
\end{figure*}

\section{Results}
\label{sec4}
\subsection{Oxygen and nitrogen abundances}
In Tables \ref{distributions_OH} and \ref{distributions_NO} we provide the overall statistics for the derived oxygen abundance and nitrogen-to-oxygen abundance ratio, respectively, in our selected sample of LINERs. We give the results for each ionizing source in the models we used to calculate the abundances for the whole sample, and we also list them according to their classification as wAGN or as retired galaxies (RG).

For 12+log(O/H), there is no difference between the median values derived for wAGN and RGs, (\ensuremath{\sim} 0.05 dex; this is lower than the grid step and compatible within the errors). It is always close to the solar value (12+log(O/H)$\sim$8.69; \citealt{Asplund_2009}). Only for the ADAF models did we obtain subsolar median values, but still no significant difference was found between the two families.  The case of  pAGB models is particularly interesting (T\ensuremath{_{eff}} = 1\ensuremath{\cdot 10^{5}} K and T\ensuremath{_{eff}} = 1.5\ensuremath{\cdot 10^{5}} K) because the difference between the resulting median values in each family increases (\ensuremath{\sim}0.1 dex), although they are still compatible within the errors. Overall, the lower values of the oxygen abundance are found in galaxies that are classified as RG, with values \ensuremath{\sim } 0.3 dex below the solar abundance. For pAGB models, both distributions reach subsolar values, below \ensuremath{\sim 0.4} dex for wAGN and \ensuremath{\sim 0.6} dex for RGs. This differentiation between AGN (and ADAF) models and pAGB models is highlighted in Fig. \ref{OH_histogram}.

Examining the log(N/O) ratio, we find a completely different picture to that depicted for oxygen. For a given particular model, log(N/O) behaves in a statistically similar way for both groups of galaxies (differences below 0.06 dex). For a given group, there is little difference between the N/O ratios obtained assuming distinct AGN models. For either pAGB or ADAF models, the median value increases with respect to AGN models (\ensuremath{\sim } 0.15 dex). Nevertheless, they are still compatible considering the standard deviations of the distributions. Overall, we found that median values cluster around the solar abundance ratio (log(N/O)$\sim$-0.86, \citealt{Asplund_2009}), but there is a wider range of values (\ensuremath{\sim } 0.9 dex) than is observed for 12+log(O/H) (\ensuremath{\sim } 0.4 dex). In Fig. \ref{NO_histogram} we show the distributions of the log(N/O) estimations for each group and for the whole sample. There is little difference among them and for each photoionization grid of models.

As shown in Fig. \ref{OH_histogram} (see also Table \ref{distributions_OH}), the oxygen abundances estimated from pAGB models with temperatures as low as \ensuremath{T_{eff} = 5\cdot10^{4}} K spread over a very narrow range of values [8.4 and 8.6], which is even shorter than the range observed from the ADAF models. The photoionization models reveal that no O\ensuremath{^{3+}} is predicted by these models  in the nebulae because their ionizing incident radiation field is comparatively softer. Hence, in order to properly provide a simultaneous estimation of the oxygen abundance (mainly traced through the sum of O\ensuremath{^{++}} and O\ensuremath{^{+}} emission lines) and the ionization parameter (mainly traced through the ratio of O\ensuremath{^{++}} and O\ensuremath{^{+}} emission lines), these models with a lower effective temperatures constrain the obtained oxygen abundance in a more restricted range than other models with higher $T_*$, which allow a larger variation in the same involved  lines, as these same models predict a higher abundance of other more highly ionized species such as O\ensuremath{^{3+}}.

\subsection{N/O versus O/H relation}
The log(N/O) versus 12+log(O/H) relation allows us to check whether the ISM follows a standard chemical enrichment history that is characterized by constant log(N/O) values at low metallicities (12+log(O/H) $<$ 8.6, \citealt[e.g.][]{Belfiore_2015}) due to the primary production of N and a positive increasing correlation at higher metallicities caused by the contribution of the secondary production in intermediate-mass stars \citep[e.g.][]{Vila-Costas_1993, Coziol_1999, Andrews_2013, Belfiore_2015, Vincenzo_2016}.

\begin{figure*}
	\centering
	\includegraphics[width=0.9\hsize]{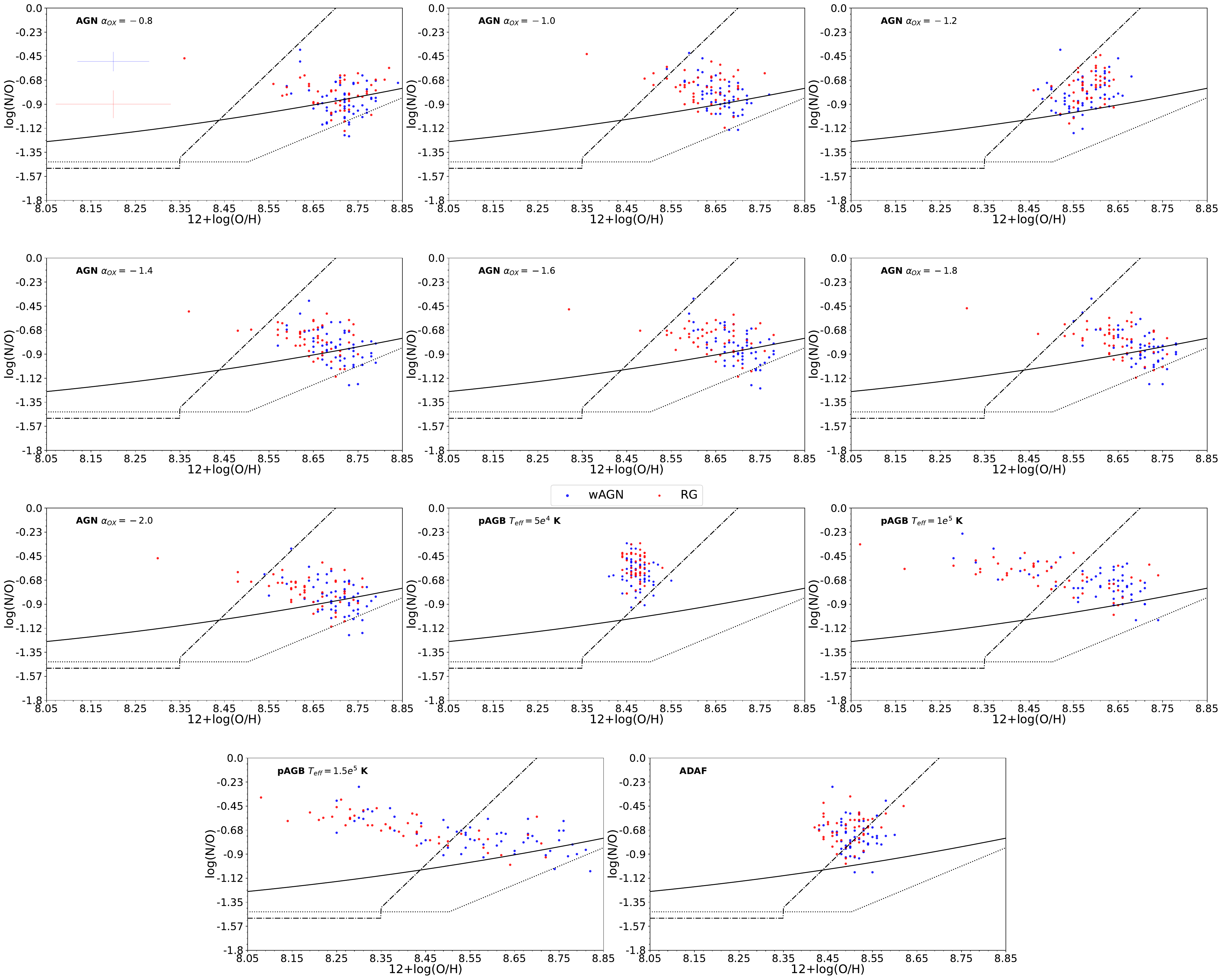}
	\caption{Relation of the nuclear estimations of log(N/O) and 12+log(O/H) in our sample of LINERs for each grid of photoionization models. The first plot shows the typical errors for the data. The solid back line represents the fit provided by \citet{Coziol_1999}, the dotted line shows the fit by \citet{Andrews_2013}, and the dash-dotted line shows the fit by \citep{Belfiore_2015}.}
	\label{NO_OH}
\end{figure*}

\begin{table*}
\caption{Results for the Pearson coefficient correlation test between log(N/O) and 12+log(O/H) (assuming that there is  no correlation as the null hypothesis.}
\label{Pearson_NO_OH}
\centering
\begin{tabular}{c | c c c } 
\hline\hline
\textbf{Model} & \textbf{Class.} & \boldmath$\rho_{Pear}$ & \textbf{p-value}  \\ \textbf{(1)} & \textbf{(2)} & \textbf{(3)} & \textbf{(4)} \\  \hline 
 & wAGN &  -0.111572 & 0.408650 \\ 
AGN $\alpha_{OX}=-0.8$ & RG &  -0.196387 & 0.180956 \\ 
 & All &  -0.175399 & 0.073503 \\ \hline
 & wAGN &  -0.526893 & 0.000025 \\ 
AGN $\alpha_{OX}=-1.0$ & RG &  -0.294983 & 0.041814 \\ 
 & All &  -0.427234 & 0.000005 \\ \hline
 & wAGN &  0.294352 & 0.026245 \\ 
AGN $\alpha_{OX}=-1.2$ & RG &  0.321981 & 0.025631 \\ 
 & All &  0.313398 & 0.001133 \\ \hline
 & wAGN &  -0.487242 & 0.000121 \\ 
AGN $\alpha_{OX}=-1.4$ & RG &  -0.419451 & 0.003001 \\ 
 & All &  -0.482328 & 0.000000 \\ \hline
 & wAGN &  -0.568648 & 0.000004 \\ 
AGN $\alpha_{OX}=-1.6$ & RG &  -0.377912 & 0.008091 \\ 
 & All &  -0.486132 & 0.000000 \\ \hline
 & wAGN &  -0.572961 & 0.000003 \\ 
AGN $\alpha_{OX}=-1.8$ & RG &  -0.474195 & 0.000662 \\ 
 & All &  -0.535446 & 0.000000 \\ \hline
 & wAGN &  -0.479811 & 0.000159 \\ 
AGN $\alpha_{OX}=-2.0$ & RG &  -0.437627 & 0.001867 \\ 
 & All &  -0.477085 & 0.000000 \\ \hline
 & wAGN &  -0.176009 & 0.190313 \\ 
pAGB $T_{eff}= 5e^{4}$ K & RG &  -0.047178 & 0.750168 \\ 
 & All &  -0.139873 & 0.154701 \\ \hline
 & wAGN &  -0.669445 & 0.000000 \\ 
pAGB $T_{eff}= 1e^{5}$ K & RG &  -0.548444 & 0.000054 \\ 
 & All &  -0.620041 & 0.000000 \\ \hline
 & wAGN &  -0.656832 & 0.000000 \\ 
pAGB $T_{eff}= 1.5e^{5}$ K & RG &  -0.757362 & 0.000000 \\ 
 & All &  -0.707616 & 0.000000 \\ \hline
 & wAGN &  -0.066046 & 0.625464 \\ 
ADAF & RG &  0.156851 & 0.287032 \\ 
 & All &  -0.006464 & 0.947821 \\ \hline
\end{tabular}
\tablefoot{For each photoionization model (1) and each group of LINERs (2), we provide the Pearson statistic (3) and the p value (4).}
\end{table*}

We show in Fig. \ref{NO_OH} the observed behavior in the N/O versus O/H diagram for our estimations of nuclear abundances in our sample of LINERs. Generally, no distinction is found between galaxies classified as wAGN or RG. While our galaxies fall in the region delimited by the scatter in the literature, there seems to be an anticorrelation for AGN models, with the exception of the estimations provided by AGN models with \ensuremath{\alpha_{OX} = -1.2}. The Pearson coefficient correlations for these models are in the range of [-0.57 to -0.11] while the p-values are lower than 0.05 in most of the cases (with the exception of AGN models with \ensuremath{\alpha_{OX} = -0.8}). This indicates that there is indeed such an anticorrelation. When pAGB models are considered (excluding those with a low effective temperature), these coefficients are even higher, \ensuremath{\sim -0.62} with pAGB models of \ensuremath{T_{eff} = 1\cdot10^{5}} K and \ensuremath{\sim -0.70} with pAGB models of \ensuremath{T_{eff} = 1.5\cdot10^{5}} K, and the p-values are even lower than 0.0005. 

The ADAF models and AGN models with \ensuremath{\alpha_{OX} = -0.8} and \ensuremath{\alpha_{OX} = -1.2} seem to follow the reported relation for SFGs, but their Pearson coefficient correlations reveal that we cannot exclude the possibility that there is no correlation at all.

\subsection{Mass-metallicity relation}
The galaxy mass assembly is tied to the enrichment of the gas-phase ISM as star formation leads to the production of metals, a fraction of which can be ejected later into the surrounding ISM in their late stages of life. The well-known mass-metallicity relation \citep[MZR][]{Lequeux_1979, Tremonti_2004, Andrews_2013, Perez-Montero_2016} is a natural result that arises from this galaxy evolution scheme. While the use of the oxygen abundance as tracer of the metallicity might lead to incorrect conclusions, as is the case when masses of inflowing and/or outflowing gas alter the ratio of oxygen and hydrogen \citep{Koppen_2005, Amorin_2010, Perez-Diaz_2024}, the same analysis can be made using the nitrogen-to-oxygen abundance ratio as a tracer of this metallicity \citep[MNOR][]{Perez-Montero_2009, Perez-Montero_2016}, with the advantage that the hydrodynamical processes that alter the MZR do not affect the MNOR.

In order to study these fundamental relations in our sample of galaxies, we retrieved the stellar mass data from the NASA-Sloan Atlas (NSA) catalog\footnote{\url{https://www.sdss4.org/dr17/manga/manga-target-selection/nsa/}.}. These values were estimated from a K-correction fit to the elliptical Petrosian fluxes assuming the initial mass function (IMF) from \citet{Chabrier_2003} and the stellar population models from \citet{Bruzual_2003}. 

As a general caveat, we used the abundance estimation from the nuclear region, whereas the reported relations were found in the chemical abundance estimation for the whole galaxy, which is more representative of the metallicity at the effective radius \citep[e.g.][]{Sanchez-Menguiano_2024} than in the nuclear region. Nevertheless, this study gives insights into whether our derived chemical abundances and the relations between them in the nuclear region can be driven by the physical galaxy parameters, which are known to play a major role in their evolution.

\begin{figure*}
	\centering
	\includegraphics[width=0.9\hsize]{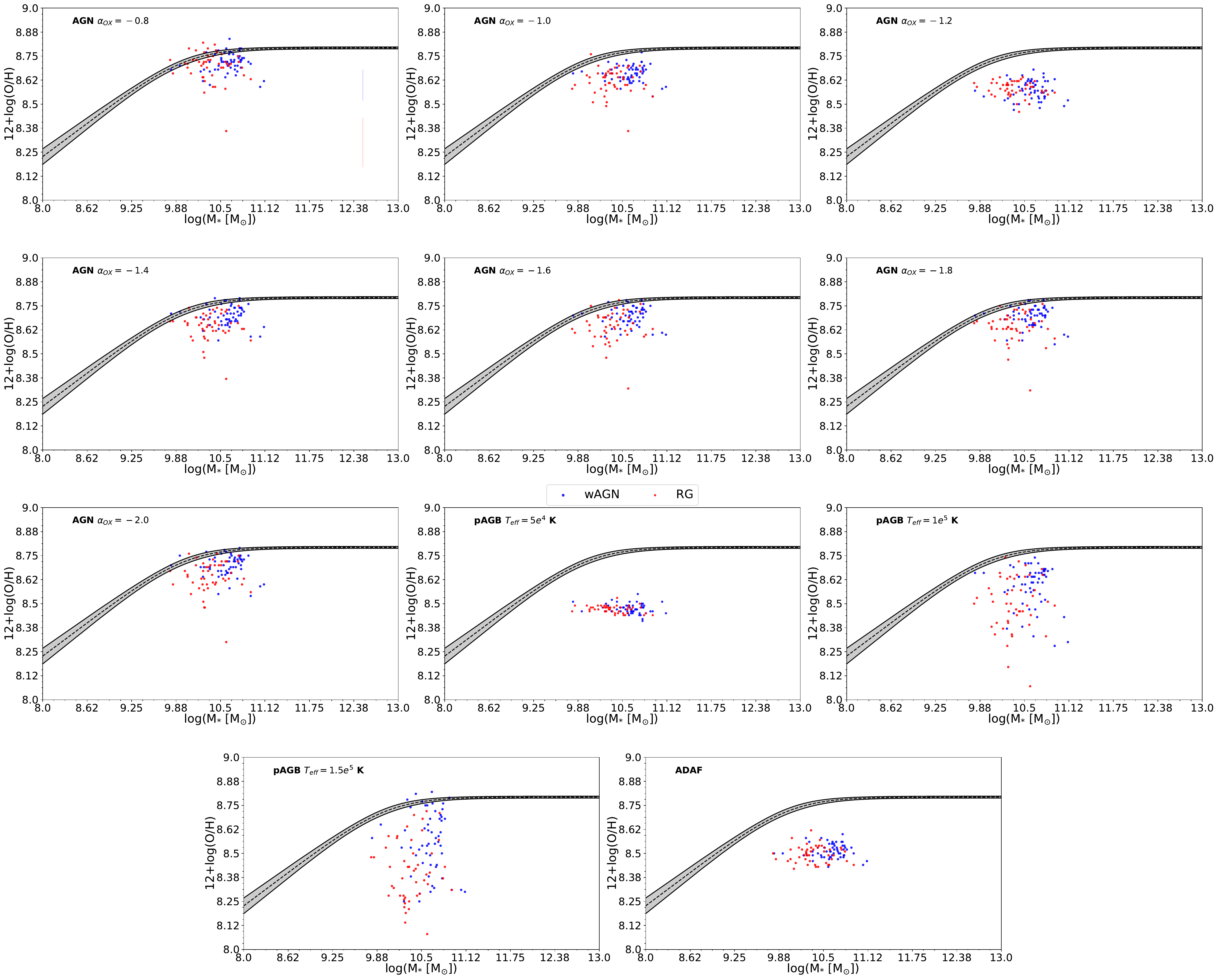}
	\caption{Mass-metallicity relation determinations in the nuclear regions of our sample of LINERs for different grids of photoionization models. The first plot shows the typical errors for the data. The dashed line represents the fit from \citet{Curti_2020}, and the gray shaded area shows the corresponding uncertainty on the relation.}
	\label{MZR}
\end{figure*}

\begin{figure*}
	\centering
	\includegraphics[width=0.9\hsize]{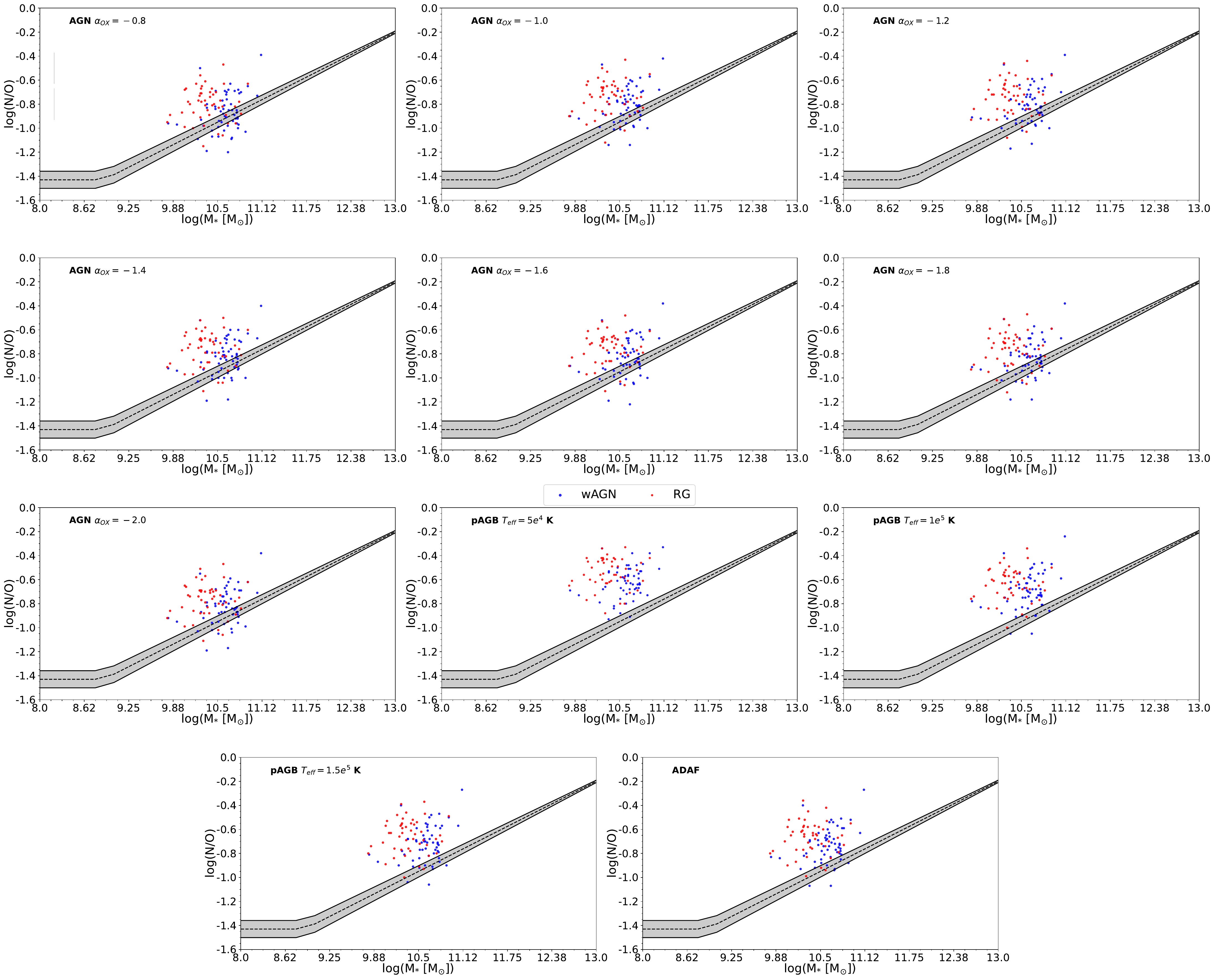}
	\caption{Mass-NO relation based on the metallicity determinations in the nuclear regions of our sample of LINERs for different grids of photoionization models. The first plot shows the typical errors for the data. The dashed line represents the fit from \citet{Andrews_2013}, and the gray shaded area shows the corresponding uncertainty on the relation.}
	\label{MNOR}
\end{figure*}

We show in Fig. \ref{MZR} the MZR in our sample of LINERs. First of all, we note that some galaxies lie on the reported relation \citep{Curti_2020} for SFGs. This is especially true for AGN models, which cluster around the relation, although they show systematically lower metallicities (\ensuremath{\sim} -0.15 dex).  Analyzing the different subtypes of galaxies, we observe that although their host galaxies are characterized by stellar masses in the same range [\ensuremath{10^{9.8} M_{\odot}}, \ensuremath{10^{11} M_{\odot}}], the median value for RGs is lower (\ensuremath{10^{10.33} M_{\odot}}) than for wAGNs (\ensuremath{10^{10.63} M_{\odot}}). For pAGB with \ensuremath{T_{eff} = 5\cdot10^{4}} K, the retrieved relation seems to lie far below the reported relation and varies little in metallicity, which again highlights the problem of assuming low effective temperatures for the ionizing source. When we considered other pAGB models, we retrieved no relation at all between mass and metallicity.

The picture that arises from the MNOR (Fig. \ref{MNOR}) complements our result form the log(N/O) versus 12+log(O/H) relation. First of all, there is very little difference in the obtained relations as a function of the assumed models. Second, most of our sample lies above the reported relation \citep{Andrews_2013}. This is consistent with the anomalous relation observed in Fig. \ref{NO_OH}: While the oxygen abundances might be consistent (or slightly lower) with their corresponding mass, the nitrogen-to-oxygen abundance ratio increases with respect to the expected ratio, which indicates that the scenario for the N evolution is more complex.
\subsection{Comparison with the extrapolation from galaxy gradients}
Although this procedure and all its associated results will be discussed in a forthcoming paper of this series, we briefly explain the procedure for estimating metallicity gradients in the same galaxies which are hosting the nuclear regions classified as LINERs. First of all, we selected the spaxels classified as HII regions in each galaxy according to the three diagnostic diagrams \citep{Baldwin_1981, Kewley_2006}. Second, we used \textsc{HCm} to  consistently derive the metallicities for each HII region using \textsc{POPSTAR} as ionizing source in the models (see Sec. \ref{sec3} for more details). Finally, for both 12+log(O/H) and log(N/O), we used a piecewise method (Tapia; Tissera private communication) to estimate the metallicity gradient for the HII regions alone, and we allowed the profile to present breaks. The gradients were calculated by normalizing distances to the \ensuremath{R_{50}} radius, defined as the azimuthally averaged SDSS-style Petrosian containing 50\% light radius from the r-band.

\begin{figure*}
	\centering
	\includegraphics[width=0.9\hsize]{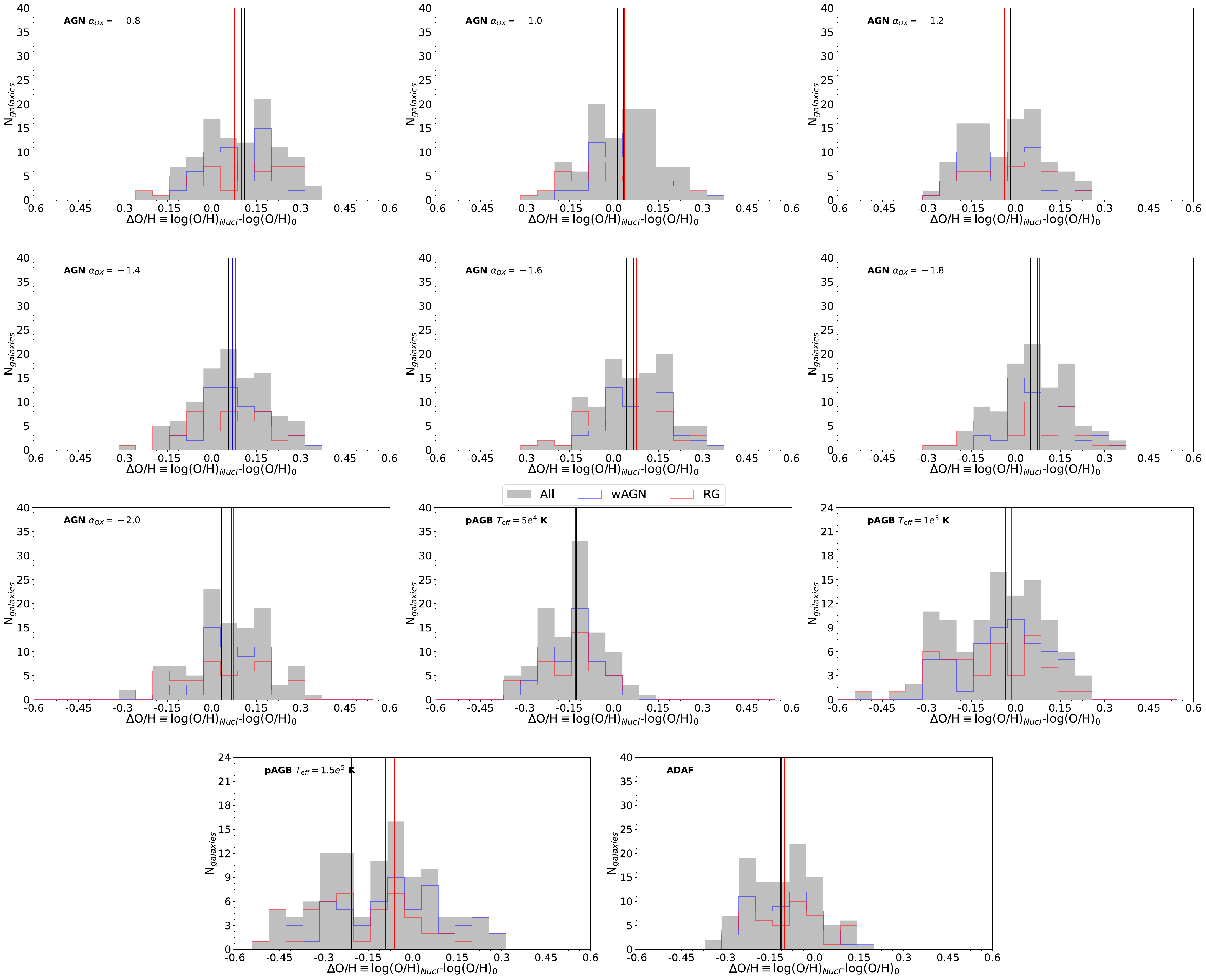}
	\caption{Histograms of the difference between 12+log(O/H)\ensuremath{_{Nucl}} as estimated in the central regions using different photoionization models and 12+log(O/H)\ensuremath{_{0}}, as extrapolated from metallicity gradients. The step-filled gray histogram corresponds to all LINERs in our sample. The step blue and red histograms correspond to wAGNs and RGs, respectively. The vertical solid lines represent the median values for each distribution.}
	\label{Gradient_estimations_OH}
\end{figure*}

\begin{figure*}
	\centering
	\includegraphics[width=0.9\hsize]{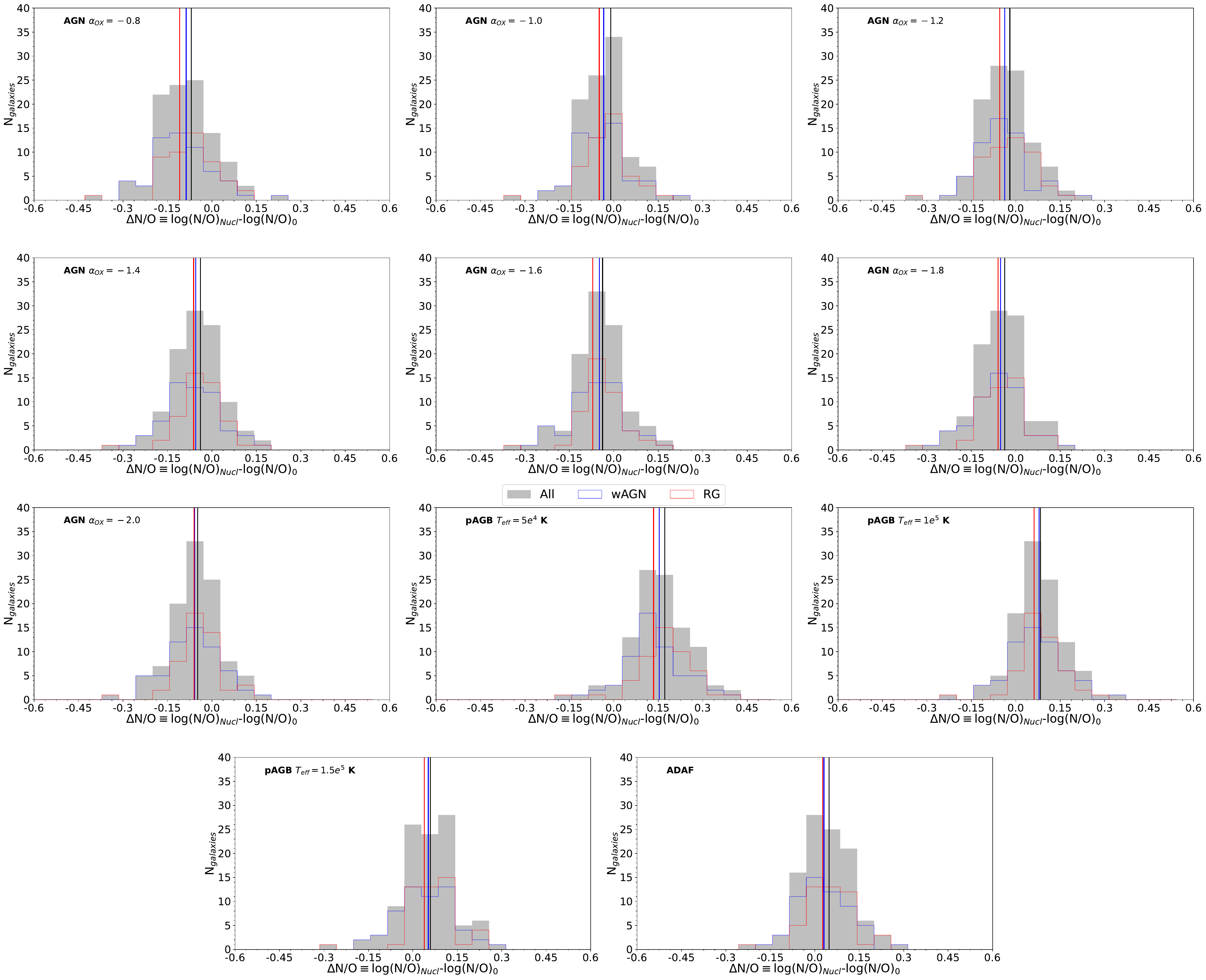}
	\caption{Same as Fig. \ref{Gradient_estimations_OH}, but for the difference log(N/O) as estimated in the central regions using different photoionization models and log(N/O)\ensuremath{_{0}} as extrapolated from metallicity gradients.}
	\label{Gradient_estimations_NO}
\end{figure*}

From Fig. \ref{Gradient_estimations_OH} we conclude that in the case of 12+log(O/H), the extrapolations from the gradients within galaxies and the estimations from the nuclear emission do not match because the distribution of the differences is wide (over 0.6 dex), even though the median difference is close to zero. We also obtained the highest offset and dispersion when the chemical abundances from the nuclear regions were estimated using pAGB models. This result holds true for both groups of galaxies we considered in this analysis  (wAGNs and RGs) . A completely different idea emerges from Fig. \ref{Gradient_estimations_NO}, where the extrapolations from the gradients and the nuclear estimations match for the log(N/O) abundance ratio. This leads to narrower distributions of the differences (\ensuremath{\sim} 0.4 dex) and to median differences much closer to zero.

\section{Discussion}
\label{sec5}
In the following section, we omit the analysis of pAGB models with \ensuremath{T_{eff} = 5\cdot10^{4}} K because we determined in Sec. \ref{sec4} in agreement with previous works \citep[e.g.][]{Krabbe_2021, Oliveria_2022} that they are inefficient in reproducing the observed emission-line ratios. Hence, throughout this discussion, we only refer to the pAGBs that are characterized by \ensuremath{T_{eff} = 1\cdot10^{5}} K and \ensuremath{T_{eff} = 1.5\cdot10^{5}} K.
\subsection{Comparison of the abundance estimations with the literature}
Although only a few studies analyzed the chemical abundances in LINERs (fewer than for SFG or Seyfert 2), we can compare our results with them. We considered the works from \citet{Annibali_2010}, \citet{Oliveria_2022, Oliveira_2024} and \citet{Perez-Diaz_2021}, as these studies analyzed statistical significant samples of LINERs.

\citet{Annibali_2010} obtained that when they used the results of \citet{Kobulnicky_1999} (which accounts for pAGN ionization) to estimate 12+log(O/H) for their sample of early-type galaxies, the abundances reported in the nuclear region lay in the range [8.20, 9.01], which agrees with the results obtained for our sample of LINERs for all models. Conversely, when they used the calibration of \citet{Storchi_1998} (which accounts for AGN activity), the range was constrained to [8.54, 8.94] because the validity range for this calibration is above 12+log(O/H) \ensuremath{>} 8.4. This range also agrees with our findings when we used AGN models as the ionizing source, although we were able to retrieve lower values than are traced by this calibration.

\citet{Oliveria_2022} analyzed a sample of 43 LINERs from MaNGA that were classified as RGs in the WHAN diagram \citep{Cid-Fernandes_2010, Cid-Fernandes_2011}. Assuming pAGB ionization, and by means of photoionization models, they used calibrations based on N2 (\ensuremath{\equiv} [\ion{N}{ii}]\ensuremath{\lambda }6584\ensuremath{\AA}/H\ensuremath{_{\alpha}}) and O3N2 (\ensuremath{\equiv} [\ion{O}{iii}]\ensuremath{\lambda }5007\ensuremath{\AA}/[\ion{N}{ii}]\ensuremath{\lambda }6584\ensuremath{\AA}) line ratios and obtained oxygen abundances in the range [8.4, 8.84]. In contrast to our results, they did not obtain any LINER with 12+log(O/H) \ensuremath{<} 8.4 as we did. In particular, in the case of RGs, we obtained that 10 RGs (21\ensuremath{\%}) show metallicities below that limit. In their nitrogen-to-oxygen abundance ratio, we found that this group of low-metallicity LINERs is also characterized by suprasolar log(N/O) ratios [-0.67, -0.34]. Since the photoionization models built by \citet{Oliveria_2022} assumed a relation of log(N/O) and 12+log(O/H), their higher abundances might be interpreted as a consequence of using nitrogen emission lines as tracers of the oxygen abundance without an independent estimation of the log(N/O) ratio. This might explain the discrepancy.

 \citet{Oliveira_2024} more recently shed more light on the oxygen abundances in LINERs. In contrast to the previous work by \citet{Oliveria_2022}, they used customized photoionization models to simultaneously constrain the N and O abundances, without assuming any relation between them, similar to our work. Their results for 40 out of the 43 previous RG LINERs on their oxygen content are in the range 8.0 \ensuremath{\leq } 12+log(O/H) \ensuremath{\leq } 9.0, which is similar to our findings, especially when pAGB models are considered for RGs (8.07 \ensuremath{\leq } 12+log(O/H) \ensuremath{\leq } 8.82). However, there is a slight discrepancy between their mean value 12+log(O/H) = 8.74 and ours for RGs 8.57 (8.50) for pAGB models with \ensuremath{T_{eff} = 1\cdot10^{5}} K (\ensuremath{T_{eff} = 1.5\cdot10^{5}} K). Even though they are still compatible within the errors, this discrepancy might be interpreted as a consequence of the fact that \citet{Oliveira_2024} considered dust-free photoionization models, whereas our grid of models accounted for a standard dust-to-gas ratio. For the nitrogen-to-oxygen abundance ratios, \citet{Oliveira_2024} reported a mean value of log(N/O) \ensuremath{=} -0.69\ensuremath{\pm}0.16 with a range [-1.05, -0.42], which agrees very well with the results we reported, although we detected only a few (three) LINERs with higher ratios. 

Finally, we also compared our results to those provided by \citet{Perez-Diaz_2021}, who analyzed 40 LINERs from the Palomar Spectroscopic Survey \citep{Ho_1993, Ho_1997} with the same method as we used, but only considered AGN models with \ensuremath{\alpha_{OX} = -0.8}. They estimated a median 12+log(O/H) = 8.63\ensuremath{\pm }0.26 and a range of metallicities given by [8.04, 8.85] for their sample of LINERs. Their median value is consistent with the values we report in this work for the different grids of photoionization models for AGN or pAGB sources and for both types of LINERs (wAGNs and RGs). However, we were not able to retrieve oxygen abundances below 12+log(O/H) $<$ 8.2 with AGN models, which might be explained by the higher spatial resolution in the Palomar Spectroscopic Survey. For log(N/O), their findings also agree with those provided by \citet{Oliveira_2024} and with ours, although the median value that we obtained for AGN models is slightly lower (log(N/O)\ensuremath{\sim }-0.80).
\subsection{Relation between the nuclear chemical abundances and the host galaxy properties}
The relation between 12+log(O/H) and log(N/O) we obtained for our sample of LINERs in MaNGA is a result that merits a thorough examination because of its connotations for the evolution of galaxies and the processes driving their chemical enrichment. Our findings are well illustrated in Fig. \ref{NO_OH} for different scenarios as a function of the SED considered in the models to calculate the abundances.
\begin{itemize}
\item \textbf{Strong negative trend}: When pAGB models were accounted for, we obtained a strong anticorrelation between log(N/O) and 12+log(O/H). When pAGB models with effective temperature \ensuremath{T_{eff} = 1\cdot10^{5}} K were considered, the Pearson correlation coefficients were -0.54 for RGs and -0.67 for wAGNs. For \ensuremath{T_{eff} = 1.5\cdot10^{5}} K, the coefficients changed to -0.75 and -0.65, respectively. In both cases, the anticorrelation reaches beyond the limits of the scatter found in the literature.
\item \textbf{Positive trend}: A positive trend was only  retrieved for an AGN model with a slope \ensuremath{\alpha_{OX} = -1.2} (Pearson correlation coefficients of 0.32 and 0.29 for RGs and wAGNs, respectively). 
\item \textbf{Slightly negative trend compatible with the scatter}: For the remaining AGN models, we found negative Pearson correlation coefficients in the range [-0.42, -0.19] for RGs and [-0.53, -0.17] for wAGNs. In all these scenarios, the position in the N/O versus O/H diagram is still compatible with the different relations reported in the literature to account for the scatter.
\item \textbf{Nonexistent trend}: Pearson correlation coefficients very close to zero (0.105 and -0.06 for RGs and wAGNs, respectively) were found when we considered the ADAF model for inefficient accretion.
\end{itemize}

In a first-order approximation, omitting dynamical processes that affect the gas-phase ISM (inflows and/or outflows), the N/O versus O/H relation should represent the two scenarios of N production: i) The production of N and O mainly from massive stars, which yields a constant log(N/O) over 12+log(O/H) $<$ 8.5 \citep{Andrews_2013, Vincenzo_2016}, and ii) an increasing log(N/O) ratio over 12+log(O/H) > 8.5 \citep{Andrews_2013, Vincenzo_2016} as a consequence of the non-negligible contribution of N via intermediate-mass stars \citep[4-7M\ensuremath{_{\odot }}][]{Kobayashi_2020} from the CNO cycles to which the oxygen already present in the stars contributes. In this scenario, the delay between nitrogen production and ejection, the N enrichment through Wolf-Rayet stars \citep[e.g.][]{Kobulnicky_1997, Lopez-Sanchez_2010} (although it might be negligible at kiloparsec scales \citep[e.g.][]{Perez-Montero_2011}) together with the differences in the star formation efficiency \citep[e.g.][]{Molla_2006} explain the scatter in the relation reported in the literature. We exemplified this here with the relations from \citet{Andrews_2013} and \citet{Belfiore_2015}. This scenario allowed us to explain not only the observed relation for AGN models in the N/O versus O/H diagram, but also the slight deviation by -0.15 dex for our sample of LINERs compared to  other reported fits of the MZR, as shown in Fig. \ref{MZR}, whereas the deviation in the MNOR diagram (see Fig. \ref{MNOR}) is stronger.

Nevertheless, the expected scatter in the O/H versus N/O relation can be even higher when the galaxy (or the central region) is not considered as a closed-box model and gas inflows and outflows as a consequence of stellar evolution or AGN activity are accounted for. As shown by inflow/outflow chemical evolution models \citep[e.g.][]{Koppen_1999} and supported by observations of interacting systems such as (ultra-) luminous infrared galaxies \citep{Perez-Diaz_2024}, the chemical enrichment of green-pea galaxies \citep[e.g.][]{Amorin_2010, Amorin_2012} and specific cases of spatially resolved galaxies such as NGC 4214 \citep{Kobulnicky_1996} or NGC 4670 \citep{Kumari_2018}. In this case, not only effects on the N production must be taken into account, but also those on the mixing and gas removal/supply, which affect the 12+log(O/H) abundance ratio. As shown by \citet{Perez-Diaz_2024}, massive infalls of metal-poor gas can drastically dilute 12+log(O/H), but log(N/O) remains mostly unaffected by the process, which might explain the high nitrogen-to-oxygen ratios (log(N/O) > -0.70) at low oxygen abundances (12+log(O/H) $<$ 8.4) observed for pAGB models. For the Teacup nebula, \citep{Villar-Martin_2024} showed that outflows might cause a pollution from the central region toward the outer parts, which would lower the oxygen abundances but maintain high N/O ratios. This result was also found by \citet{Oliveira_2024} for customized photoionization models based on pAGB SEDs in their sample of RGs. If this is the scenario, then we should expect a significant scatter in the MZR diagram, and the MNOR diagram should mimic the behavior for the remaining models. This is what we obtained for pAGB models.

\begin{figure*}
	\centering
	\includegraphics[width=0.95\hsize]{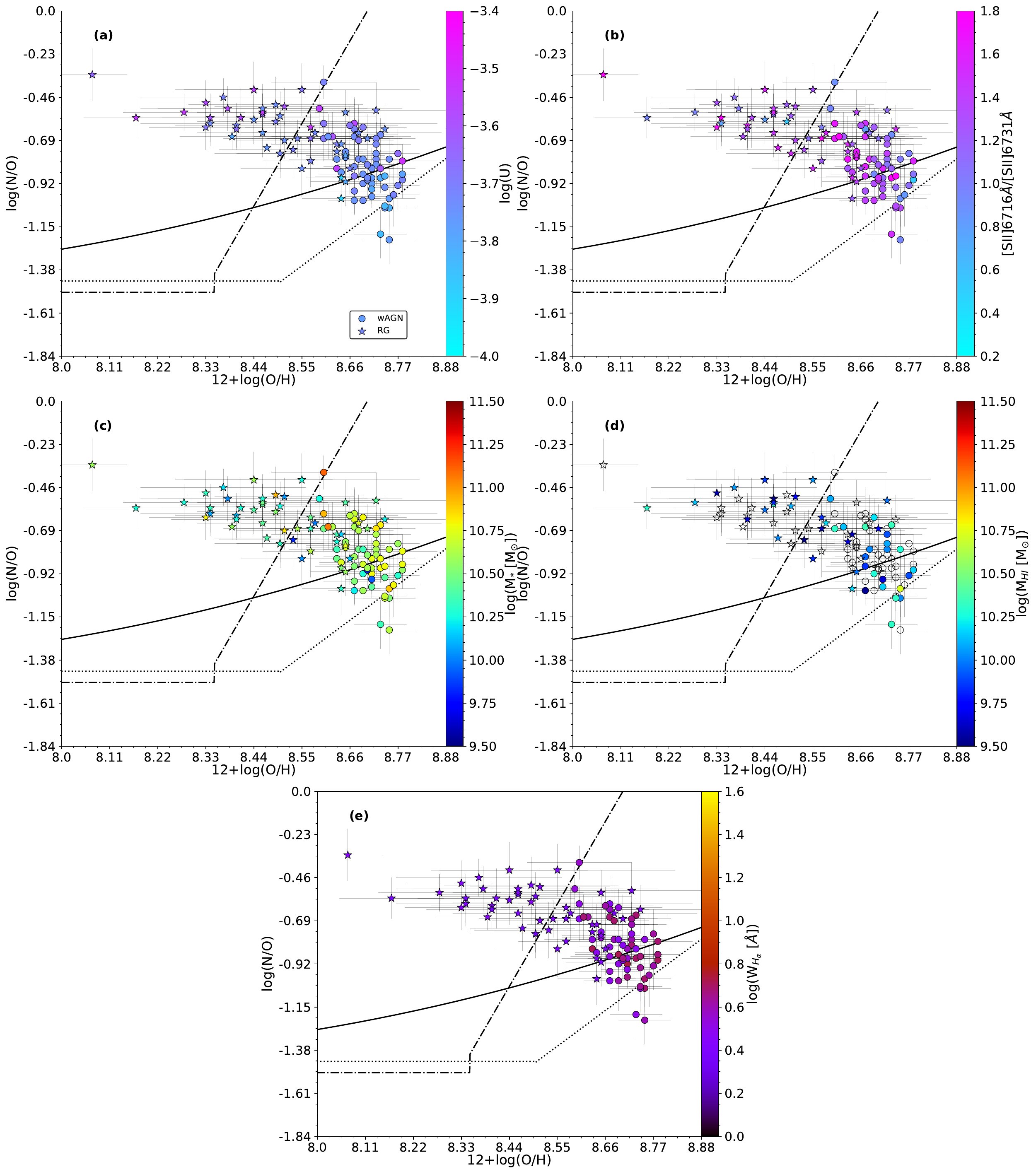}
	\caption{log(N/O) vs. 12+log(O/H) diagram for our sample of LINERs. The dots represent galaxies classified as wAGN, and the chemical abundances correspond to AGN models with \ensuremath{\alpha_{OX} = -1.6}. The stars correspond to galaxies classified as RG whose chemical abundances were estimated assuming pAGB models with \ensuremath{T_{eff} = 1\cdot10^{5}} K. The color bar shows different properties: a) The ionization parameter as estimated from emission lines. b) The sulfur ratio tracer of the electron density. c) The stellar mass as retrieved from the NSA catalog. d) The HI mass as retrieved from the NSA catalog. e) The equivalent width of H\ensuremath{_{\alpha}}. The empty symbols correspond to galaxies without a measurement of the represented property.}
	\label{NO_OH_properties}
\end{figure*}

In order to assess the origin of the observed anticorrelation between O/H and N/O under certain model assumptions, we explored the influence of different galaxy and gas properties, as shown in Fig. \ref{NO_OH_properties}. For this plot, we took estimates for wAGN galaxies assuming AGN models into account and pAGB models (\ensuremath{T_{eff}= 1\cdot10^{5}} K) for RG galaxies. First of all, it is clearly shown that whereas wAGNs follow the N/O versus O/H relation within the scatter, this is not the case for RGs because they show an anticorrelation that exceeds the scatter. In panels a) and b) we analyze whether this relation is driven by physical properties of the gas-phase ISM, such as the electron density or the ionization parameter. None of them are correlated. In panel c) we show the anticorrelation as a function of the stellar mass. There is again no relation at all. In panel d) we show the HI mass reported for each galaxy. This panel allows us two conclusions: i) neither the scatter within the reported relation nor the anticorrelation are driven by the amount of gas, as is observed for the sample of wAGNs and, ii) only one of the LINERs classified as RG and with high log(N/O) (>-0.6) and low 12+log(O/H) ($<$ 8.25) has a slightly higher gas content than the rest of the sample. However, because we lack measurements for most of them, it is not possible to asses whether this object is an outlier. Finally, we present in panel e) the equivalent width of H\ensuremath{_{\alpha}}. Based on these results, we can explore the inflow and/or outflow scenarios for the anticorrelation.

For the inflow of gas as the major driver of the observed relation, we found several caveats. The first caveat is the origin of the gas. In the case of green-pea galaxies, the impact of metal-poor high-velocity clouds \citep[HVCs, e.g.][]{Koppen_2005} or gas accretion from the cosmic web \citep{Sanchez-Almeida_2015} was proposed. In the case reported by \citet{Perez-Diaz_2024}, gas is driven by merger interaction, and the amount of gas that dilutes the metallicity from 12+log(O/H) = 8.7 (solar) to 12+log(O/H) = 8.1 can be expected to arise from the reservoirs of gas from the galaxy itself as well as tidal gas taken from the companion \citep[e.g.][]{Montuori_2010, Rupke_2010, Sparre_2022}. Considering that the anticorrelation holds in RGs, which are assumed to be populated by pAGB, it is necessary to justify not only the presence of this gas (which is not found in the mass of HI), but also the physical driver toward the nuclear region of these galaxies. The second caveat is also related to the first, and it is the chemical composition of the inflowing gas. To dilute the oxygen content by more than 0.5 dex, metal-poor gas is required, and because gas flows from the outer to the inner parts, an almost flattened gradient is expected. When the timescale of the gas dynamics is long enough, a positive gradient from the center to the outer parts is expected. This scenario and its implications are explored in the second paper of this series. The third caveat is the consequences of this inflowing gas. Due to the amount of gas required, and as reported by \citet{Perez-Diaz_2024} in (U)LIRGs as well as in simulations \citep{Montuori_2010}, an increase in the star formation rate is expected, and the equivalent width of H\ensuremath{_{\alpha}} should be higher \citep{Cid-Fernandes_2010} for the most extreme cases. This implies that they would have never been classified as RGs. Moreover, there is no trend of an increasing  equivalent width of H\ensuremath{_{\alpha}}, as shown in Fig. \ref{NO_OH_properties} (panel e).

On the other hand, when we consider the outflow scenario as the major driver of this relation, some caveats arise. The first caveat is that the observational results for a sample of LINERs classified as RGs presented by \citet{Oliveira_2024} show no evidence of disturbed kinematics in the gas, as would be expected from the outflow scenario. Another caveat arises from the metallicity gradient: An outflow would carry gas for some distance, depending on its power. A change in the metallicity gradient for both 12+log(O/H) and log(N/O) is therefore expected, with a change in the slope of the gradient at some point as a consequence of the additional enrichment, as is the case in AGN-powered outflows \citep[e.g.][]{Villar-Martin_2024}. Although we will explore this scenario in the next paper, we already have some hints. If outflows pollute the ISM at galactic scales, then the increase in log(N/O) would translate into an overestimation of the log(N/O) extrapolated to the nuclear region. In contrast to this idea, we find in Fig. \ref{Gradient_estimations_NO} that the nuclear estimation and nuclear extrapolation agree very well. Finally, if outflows are indeed present in sources that are assumed to be ionized by pAGB, the most likely scenario is that these systems of pAGB stars drive the outflows. Because they are required to ionize the whole nuclear region, the gas would account for an additional contribution of shock emission. We therefore assume that purely pAGB models are not suitable for modeling the ionizing structure.

Although we cannot entirely exclude an inflowing and/or outflowing scenario that would explain the observed anticorrelation between 12+log(O/H) and log(N/O) when pAGB stars are assumed to cause the emission on some LINERs, we must bear in mind the numerous caveats that these scenarios imply and that should be revised in future works. We also highlight that this is not a problem when all LINERs are considered to host AGNs because the anticorrelation is much weaker and compatible with the scatter in the relation, as reported in the literature.
\subsection{Source of ionization}
The nature of the source of the gas ionization in LINER-like galaxies is still an open question \citep[e.g.][]{Binette_1994, Kewley_2006, Nemmen_2014}. Of the different scenarios that were proposed to explore the origin of ionization of the nuclear gas-phase ISM in LINERs, we explored three scenarios: i) Standard AGN models with different SEDs shapes, ii) pAGB stars, and iii) inefficient accretion leading to an ADAF regime. With the exception of shocks, which were omitted from this study due to the large number of free parameters required to model this emission \citep[e.g.][]{Sutherland_2017}, all of them represent the scenarios proposed in the literature to explain this emission.

\begin{figure*}
	\centering
	\includegraphics[width=0.9\hsize]{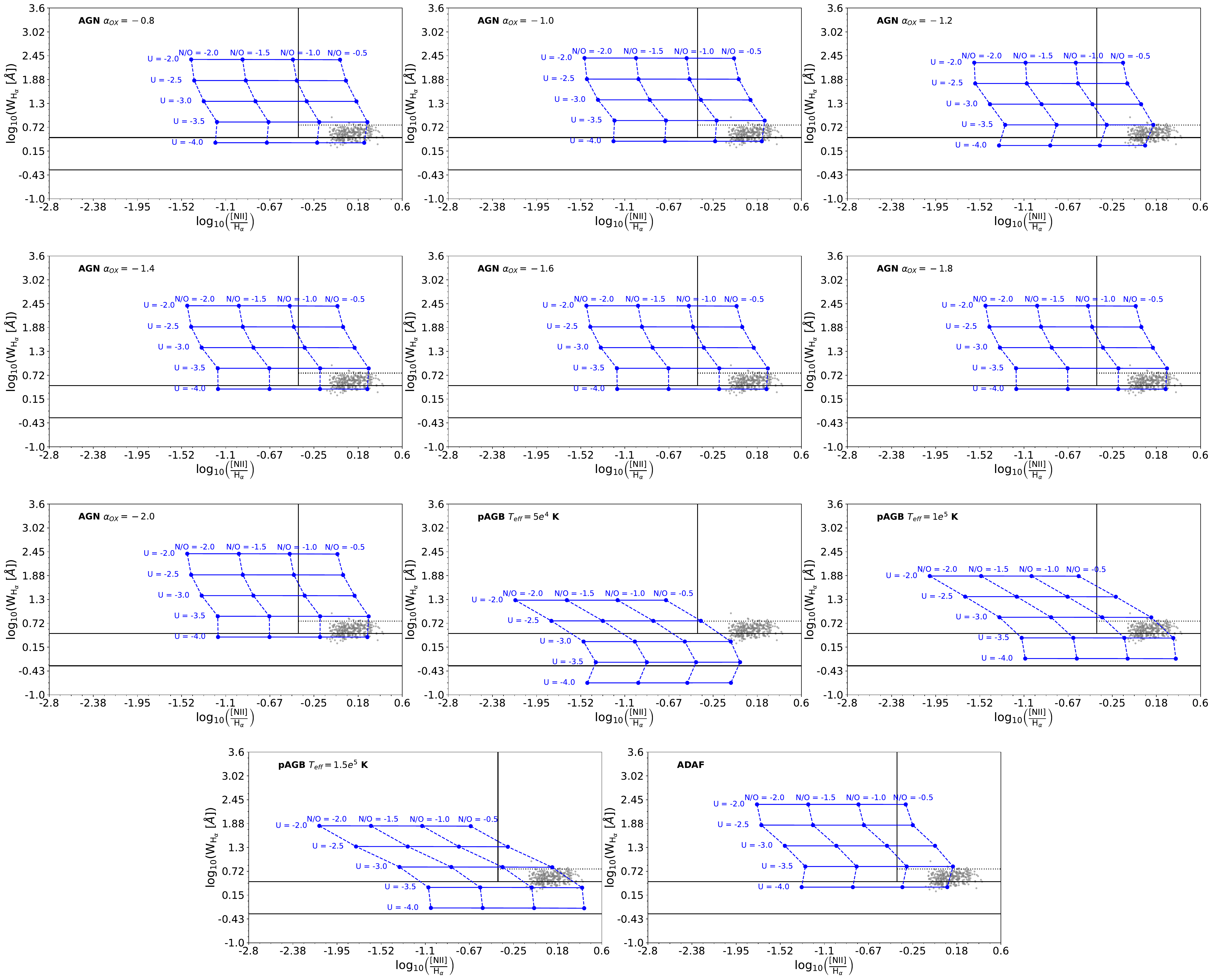}
	\caption{WHAN diagram for the LINER-like nuclear regions in our sample (gray dots) showing the coverage of the different grids of photoionization models. The grid of models was obtained by limiting the oxygen abundance to 12+log(O/H)=8.6 and employing different values of the ionization parameter log(U) as well as the nitrogen-to-oxygen abundance ratios log(N/O).}
	\label{whan_models}
\end{figure*}

One of the major caveats of the identification of LINERs based on their optical emission lines arises from its possible classification from the diagnostic diagrams. Due to the uncertainty in the classification through the BPT diagrams \citep{Baldwin_1981, Kauffmann_2003, Kewley_2006}, \citet{Cid-Fernandes_2010} proposed an alternative version using the information from the equivalent width of H\ensuremath{_{\alpha }} to distinguish between AGNs and retired galaxies. To demonstrate the caveats of distinguishing wAGNs and RGs, we built a semi-empirical galaxy grid by rescaling the W\ensuremath{_{H_{\alpha}}} predictions from our grids of photoionization models to the observed Balmer lines and continua in the central spaxels. We then consistently retrieved the expected position on the WHAN diagram for our semi-empirical modeled galaxies based on different properties of the gas-phase ISM (log(N/O) and log(U)). Fig. \ref{whan_models} shows that the position of a galaxy in the WHAN diagram is better provided by a particular combination of log(N/O) (which allows movement in the X-axis) and log(U) (which allows changes in the equivalent width) rather than by the ionizing source, although there are differences between the different ionizing sources, but they are not so extreme as to provide a clear classification.

Our study of the chemical abundances in the gas-phase ISM in our selected sample revealed some interesting constraints when different SEDs based on these assumptions were considered in the models. First of all, pAGB models are required to have high effective temperatures (\ensuremath{T_{eff} \geq 1\cdot10{5}} K) to reproduce the observed properties \citep{Krabbe_2021} and to obtain reasonable estimations of the oxygen abundance that are consistent with other galactic properties, as found in this work. ADAF models simulate an ionizing SED that is produced by inefficient accretion \citep[e.g.][]{Nemmen_2014} and are not valid for determining LINERs out of the range of 8.4 $<$ 12+log(O/H) $<$ 8.6, in contrast to the rest of models explored in our work and in previous studies \citep{Perez-Diaz_2021, Oliveria_2022, Oliveira_2024}.

The remaining two scenarios, pAGB stars or AGN standard activity, show slight differences in the estimated oxygen content. The first scenario provides wider ranges and a lower median value (12+log(O/H) = 8.50-8.57) than the second (12+log(O/H) = 8.68-8.72), although the differences are still compatible within the dispersion observed in our sample. As our sample is according to the WHAN diagram composed of wAGNs and RGs in a similar proportion (54.3\% and 45.7\%, respectively), it might be expected that the pAGB scenario is favored in RGs and the AGN is more accurate for wAGNs. However, we found no such evidence. Conversely, based on the position of RGs in the MZR (see Fig. \ref{MZR}), when pAGB stars are the source of ionization, then there is an offset of more than 0.5 dex from the expected value of 12+log(O/H) according to their stellar mass. Moreover, when we account for the secular evolution within the galaxy, we obtain a higher offset between the oxygen abundance as calculated assuming pAGB models and the values extrapolated from the gradient, even in the case of LINERs classified as RGs.

For the pAGB scenario, an additional problem arose. When we considered that pAGB stars emit \ensuremath{5\cdot10^{46}} ionizing photons per second on average \citep{Valluri_1996}, and the average H\ensuremath{_{\alpha}} luminosity in the integrated nuclear (r = 1kpc) region is log(L(H\ensuremath{_{\alpha}}) [erg/s]) = 39.35 \citep[e.g.][]{Krabbe_2021}, we can make use of the relation between the number of ionizing photons and the expected H\ensuremath{\alpha } luminosity \citep{Osterbrock_book}, which is given by
\begin{equation}
\label{n_phtons} N_{ion. \ ph} [s^{-1}] \approx 7.31\cdot10^{11} L \left( H_{\alpha } \right) [erg\cdot s^{-1}].
\end{equation}
to estimate the number of pAGB stars needed to reproduced that central emission assuming ionization-bounded conditions,
\begin{equation}
\label{n_stars} N_{pAGB} = \frac{N_{ion. \ ph, tot}}{N_{ion. \ ph, pAGB}} \approx 1.463\cdot10^{-35} L \left( H_{\alpha } \right) [erg\cdot s^{-1}],
\end{equation}
which in this particular case yields \ensuremath{3.2\cdot10^{4}} stars within a radius of 1 kpc. Since all these stars ionize the gas nowadays today, we can assume that they were formed roughly at the same time. Assuming the initial mass function (IMF) form from \citet{Salpeter_1955}, we can then obtain the expected number of stars in a range of mass as
\begin{equation}
\label{alpeter} N = \int_{M_{min}}^{M_{max}}  \xi \left( M \right) dM = \int_{M_{min}}^{M_{max}} \xi_{0} M^{-2.35} dM.
\end{equation}
We consider that the expected mass of the progenitors of pAGB stars ranges in [1.5M\ensuremath{_{\odot }}, 8M\ensuremath{_{\odot }}] \citep[e.g.][]{Ventura_2017}. Given that the validity of the Salpeter IMF has a minimum value of 0.5M\ensuremath{_{\odot }} and that we only consider an upper limit of 20M\ensuremath{_{\odot }} (quite conservative), the probability of finding one pAGB star can then be calculated as
\begin{equation}
\label{prob} p = \frac{\int_{1.5M_{\odot }}^{8M_{\odot }} \xi_{0} M^{-2.35} dM}{\int_{0.5M_{\odot }}^{20M_{\odot }} \xi_{0} M^{-2.35} dM},
\end{equation}
which gives 20.46\%, which is consistent with the idea that per globular cluster (\ensuremath{\sim 10^{5}} L\ensuremath{\odot }) one pAGB star is expected \citep{Renzini_1986, Valluri_1996}. This is an upper limit to the probability of finding one pAGB star. Thus, the minimum number of globular clusters expected within a radius of 1 kpc should be around \ensuremath{\sim 3.2\cdot10^{4}}. For comparison, the number of globular clusters reported in the inner part of the Milky Way (r $<$ 3.5 kpc) is lower than 100 \citep{Bica_2024}. Thus, a pAGB nature of the ionizing source is less favored. 

Overall, we cannot conclude on the source of ionization without other spectral regimes. For the AGN scenario, X-ray counterparts from the accretion disk surrounding the SMBH must be expected. This approach was followed by \citet{Perez-Diaz_2021} to justify the assumption of AGN activity when they analyzed their sample of LINERs from the Palomar Spectroscopic Survey. Additional constrains can be obtained with X-ray data, such as, the shape (\ensuremath{\alpha_{OX}}) of the SED. Another important constraint can be found in the use of infrared (IR) emission lines, which trace highly ionic species such as O$^{++}$, O$^{3+}$, Ne$^{++}$,  Ne$^{4+}$, Ne$^{5+}$, Ar$^{++}$, Ar$^{4+}$, or Ar$^{5+}$ \citep{Perez-Diaz_2022, Perez-Diaz_2024b}. They might provide SED constraints by means of diagnostic diagrams such as the softness diagram \citep[e.g.][]{Perez-Montero_2024}. We cannot exclude the possibility that pAGB stars (\ensuremath{T_{eff} \geq 1\cdot10^{5}} K) contribute to the ionization of the gas-phase ISM together with central AGN emission, although the exact combination of these two ionizing  sources is a much more complex issue for which better constraints are needed.
\subsection{N/O as an unbiased tracer of chemical enrichment}
Throughout this study, we distinguished between LINERs classified as wAGNs and RGs as a consequence of the uncertainty in the nature of the ionizing source in these objects \citep[e.g.][]{Marquez_2017}. For this reason, we explored different ionizing scenarios. On the one hand, we obtained different 12+log(O/H) estimations when we accounted for the AGN or the pAGB scenario (see Table \ref{distributions_OH} and Fig. \ref{Gradient_estimations_OH}). In contrast, an almost negligible change (considering the uncertainties) was found for the estimation of log(N/O) for all the assumed SEDs.

The importance of estimating log(N/O) was already highlighted by several authors \citep[e.g.][]{Perez-Montero_2009, Vincenzo_2016, Perez-Diaz_2022}. First of all, its importance relies on the use of estimators of the oxygen abundance based on the N emission lines, as an independent constraint on log(N/O) is needed to avoid effects of enhancing the metallicity through the high log(N/O) ratios (see, e.g., the difference between \citealt{Oliveria_2022} and \citealt{Oliveira_2024}). Second, an estimation of N/O  complements not only the information, but also the degree of the chemical enrichment \citep[e.g.][]{Pilyugin_2004, Perez-Montero_2013, Perez-Montero_2016, Vincenzo_2016}. Third, it remains mostly unaffected by gas dynamics that can alter the chemical composition as traced by 12+log(O/H) \citep[e.g.][]{Edmunds_1990, Amorin_2010, Amorin_2012, Koppen_2005, Sanchez-Almeida_2015, Perez-Diaz_2024}. 

In addition, based on this study of the use of optical emission lines for estimating chemical abundances, we can also add another advantage to the use of log(N/O). It can be properly estimated without a bias when a particular ionizing source is assumed. Regardless of the ionizing source assumed for the gas-phase ISM, the log(N/O) abundances in our sample of LINERs are in the range [-1.20, -0.27]. They cluster around the value of \ensuremath{\sim} -0.79 (slightly suprasolar).
\section{Conclusions}
\label{sec6}
We analyzed a sample of 105 optically selected LINERs from SDSS-IV MaNGA. In particular, we studied the chemical abundances using photoionization models in their nuclear region that accounted for different scenarios that represent the uncertainty in the source of ionization: AGN models with different shapes, pAGB models with different effective temperatures, and inefficient-accretion AGN models in which the accretion disk was truncated (ADAF). To assess whether one or multiple scenarios might be feasible, we also used the WHAN diagram to distinguish between galaxies with intermediate equivalent widths of H\ensuremath{\alpha }, which are expected to be weak AGNs (wAGNs), and galaxies with low equivalent widths, which are though to be retired galaxies (RGs) and are thus powered by hot, old stellar populations.

Our results showed that oxygen abundances (12+log(O/H)) in the nuclear region of LINERs, with independent subtypes (wAGN or RG) and the SED, are spread over a wide range of values: 8.08 \ensuremath{\leq } 12+log(O/H) \ensuremath{\leq } 8.82 in the pAGB scenario, and 8.30 \ensuremath{\leq } 12+log(O/H) \ensuremath{\leq } 8.84 for the AGN scenario. The corresponding median value is 12+log(O/H )= 8.69 (solar) for the AGN scenario, and it drops to 12+log(O/H) = 8.53 (slightly subsolar) in the pAGB scenario. The nitrogen-to-oxygen abundance ratios (log(N/O)) are mainly suprasolar, within the range -1.20 \ensuremath{\leq } log(N/O) \ensuremath{\leq } -0.38 (when AGNs are considered) or -1.07 \ensuremath{\leq } log(N/O) \ensuremath{\leq } -0.27. The median value reported in the pAGB scenario is slightly higher (log(N/O) = -0.69) than in the AGN scenario (log(N/O) = -0.79), and in both cases, it is slightly suprasolar. Overall, LINERs are mainly characterized by suprasolar log(N/O) ratios and sub- or close to solar 12+log(O/H) ratios. When we analyzed the behavior of these nuclear estimations with host galaxy physical properties such as stellar mass, we found no correlation, and we report a scatter in the MNOR diagram and not in the MZR diagram.

The assumptions on the nature of the ionizing source are critical for the 12+log(O/H), since pAGB models introduce much more scatter in the MZR relation and in their comparison with the extrapolation from metallicity gradients. This is highlighted in the behavior of the log(N/O) versus 12+log(O/H). When LINERs are considered to be powered by AGNs, little correlation is found between the two quantities, but the data fit the scatter of the reported relation in the literature. In contrast, when LINERs are considered to be powered by pAGB stars, then a negative correlation is found between the two quantities, as previously reported in the literature. We discussed several scenarios to explain this anticorrelation, although more specific studies of this result must be performed in order to obtain more robust conclusions.

Although we cannot rule out the possibility that emission in LINERs might be explained by pAGB stars, our study revealed that several problems arise from this assumption. First of all, in order to reproduce the emission detected in the nuclear region, a large number of pAGB stars is required, and it is hard to explain both the number and the simultaneous coexistence of those stars. Second, the results for the chemical enrichment (as seen from the log(N/O) vs. 12+log(O/H) diagram) indicate a complex scenario in which extreme events are needed to explain the  observed relation. In contrast, when LINERs are assumed to be powered by AGNs, the chemical enrichment scenario does not depart from the reported trends in the literature in general, although some indications of an anticorrelation were found that might be explained with the scatter caused by the delay in the N production as well as other gas dynamical effects on 12+log(O/H), but these are less extreme than in the pAGB scenario.

\section*{Data availability}
Tables B1--B5\ref{Sample_ionization} are available in electronic form at the CDS via anonymous ftp to cdsarc.u-strasbg.fr (130.79.128.5) or via http://cdsweb.u-strasbg.fr/cgi-bin/qcat?J/A+A/.

\begin{acknowledgements}
We acknowledge support from the Spanish MINECO grant PID2022-136598NB-C32. We acknowledge financial support from the Severo Ochoa grant CEX2021-001131-S MICIU/AEI/10.13039/501100011033. This research made use of \textsc{Astropy}, which is a community-developed core Python package for Astronomy \citep{Astropy_2013, Astropy_2018, Astropy_2022}, and other software and packages: \textsc{Numpy} \citep{Walt_2011}, and \textsc{Scipy} \citep{Virtanen_2020}.
The plots for this research were created using \textsc{Matplotlib} \citep{Hunter_2007}. We acknowledge the fruitful discussions with our research team. We thank the anonymous referee for the constructive report that improved this manuscript. E.P.M. acknowledges the assistance from his guide dog, Rocko, without whose daily help this work would have been much more difficult.
\end{acknowledgements}

\bibliographystyle{bibtex/aa} 
\bibliography{hcm}

\appendix
\section{Comparison of estimations of chemical abundances}
We present in this appendix a global comparison among the chemical abundances estimated in the nuclear region of our sample of LINER-like galaxies based on different assumptions for the ionizing source.

\begin{figure*}
	\centering
	\includegraphics[width=0.99\hsize]{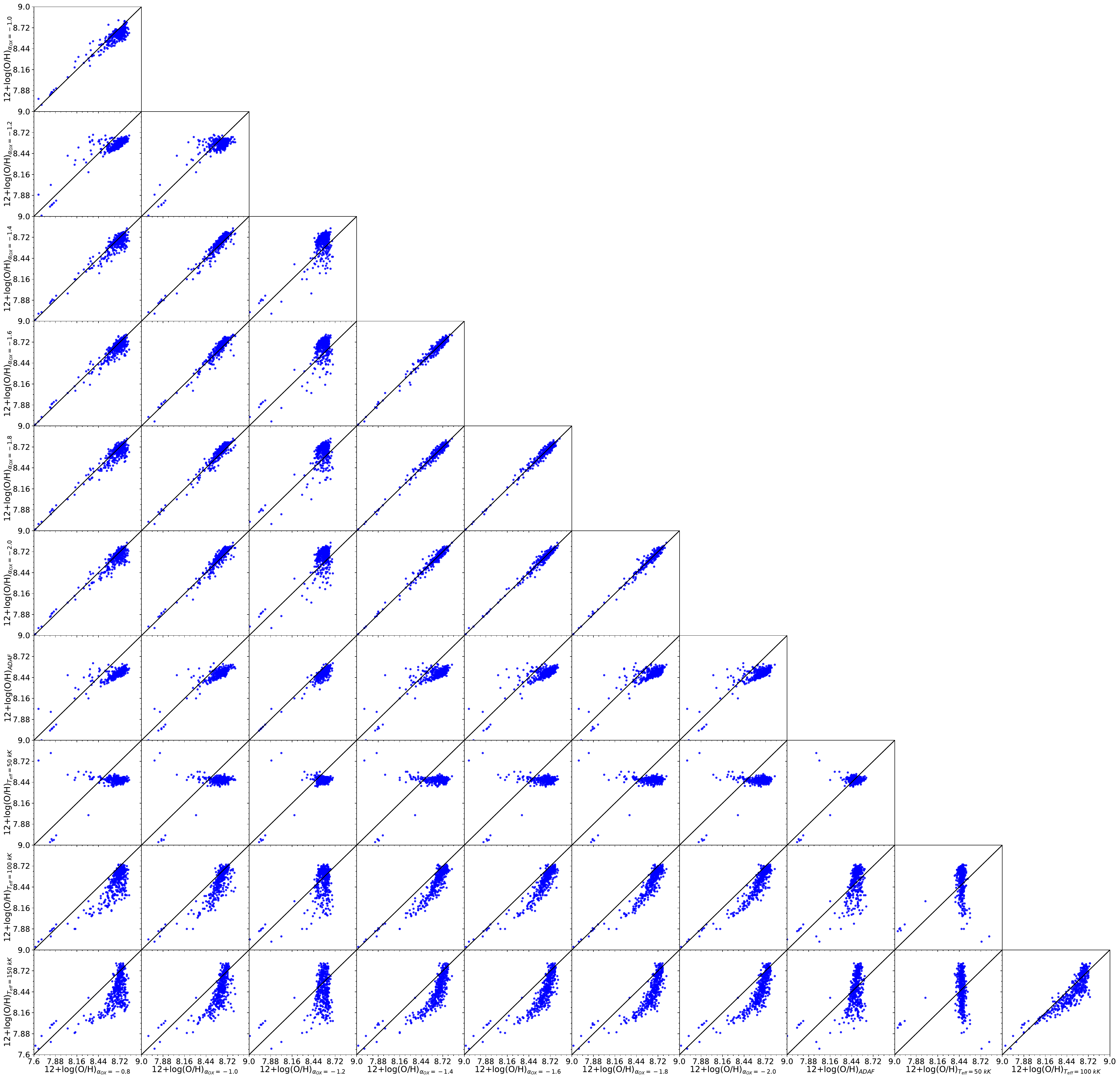}
	\caption{Comparative plot of oxygen abundances in the nuclear regions of our sample of LINERs for each grid of photoionization models considered in this work.}
	\label{comp_all_OH}
\end{figure*}

We show in Fig. \ref{comp_all_OH} the comparison among the oxygen abundances estimated in the nuclear region of our sample of LINER-like galaxies based on different grids of photoionization models. Despite there is a little discrepancy when AGN models with \ensuremath{\alpha_{OX} = -1.2} are considered, we can conclude that changes in the slope (\ensuremath{\alpha_{OX}}) of the AGN SED do not induce changes in the estimation of 12+log(O/H). In the same way, very little change is found when switching from pAGB models characterized by \ensuremath{T_{eff}= 1\cdot10^{5}} K to \ensuremath{T_{eff}= 1.5\cdot10^{5}} K. When comparing AGN models and pAGB models, we observed that from 12+log(O/H) = 8.1 to 12+log(O/H) = 8.7 (solar value), pAGB models predict systematically lower chemical abundances than AGN models. Fig. \ref{comp_all_OH} also highlights the problem of assuming pAGB models with low effective temperatures (\ensuremath{T_{eff}= 5\cdot10^{4}} K).

\begin{figure*}
	\centering
	\includegraphics[width=0.99\hsize]{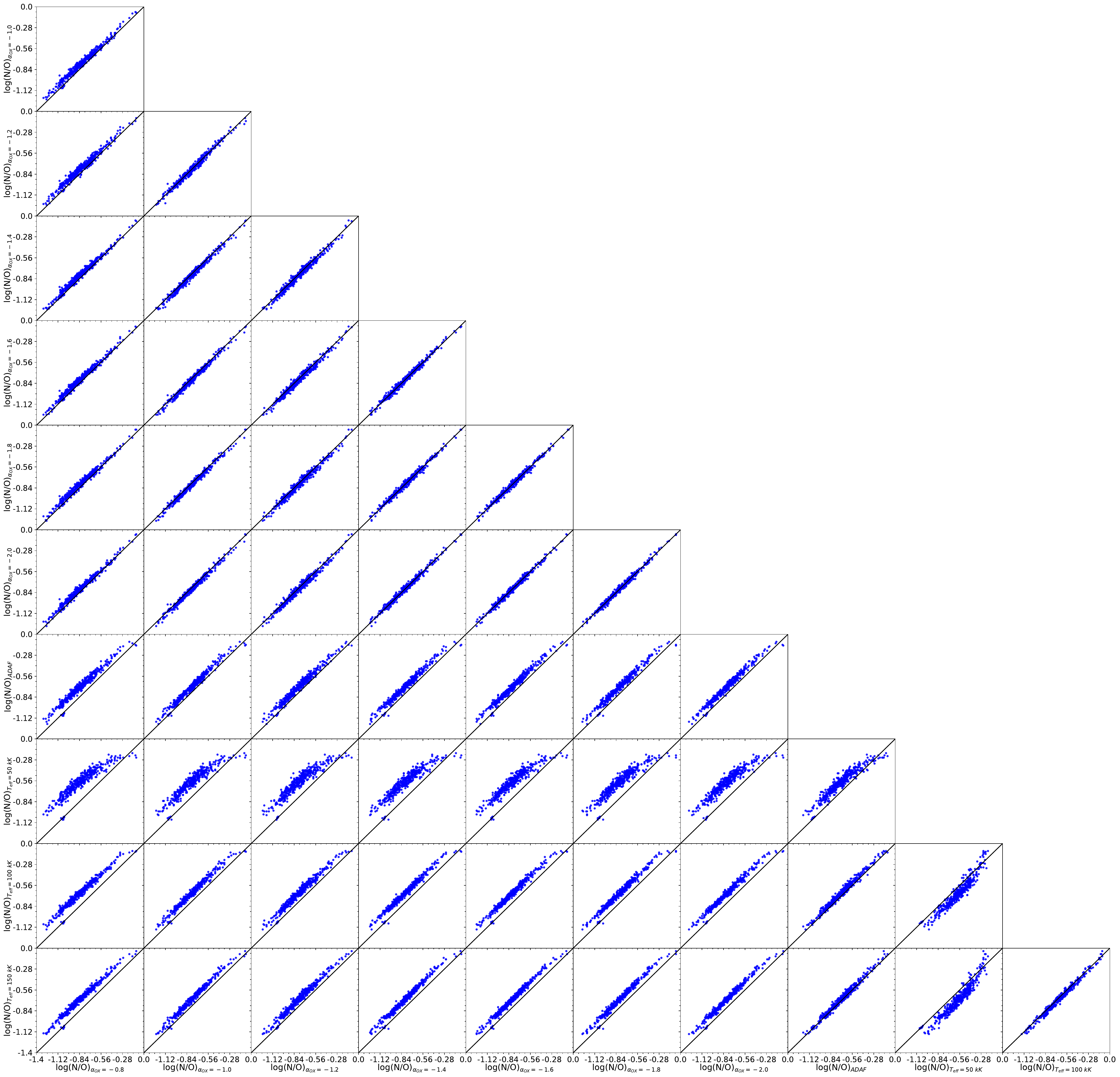}
	\caption{Same as Fig. \ref{comp_all_OH} but for the nitrogen-to-oxygen abundance ratio.}
	\label{comp_all_NO}
\end{figure*}

On the contrary, Fig. \ref{comp_all_NO} shows the excellent agreement among the estimations of log(N/O). Omitting pAGB models characterized by \ensuremath{T_{eff}= 5\cdot10^{4}} K, which have been already probed to be inappropriate for this analysis, there is correlation one-to-one correlation for all photoionization models considered. We also noticed that there is a small offset between AGN models and ADAF or pAGB models, although the correlation one-to-one still holds.

\section{Data}
We present in this appendix the full dataset used for this work. Table \ref{Sample_selection} contains all ancillary data for our sample of galaxies. Table \ref{Sample_lines} lists all optical spectroscopic properties of the nuclear regions of our sample of LINER-like galaxies. Tables \ref{Sample_oxygen} - \ref{Sample_ionization} list all the estimations of chemical abundances and ionization parameters for the nuclear regions of our sample.

\begin{table*}
\caption{List of galaxies and host galaxy properties in our sample of LINER-like galaxies.}
\label{Sample_selection}
\centering
\begin{tabular}{c c c c c c c} 
\hline\hline
\textbf{MaNGA ID} & \textbf{z} & \textbf{Class. BPT} & \textbf{Class. WHAN} & \boldmath$R_{50}$ \textbf{[arcsec]}  & \boldmath$\log M_{*} [M_{\odot}]$ & \boldmath$\log M_{HI}  [M_{\odot}]$\\ \textbf{(1)} & \textbf{(2)} & \textbf{(3)} & \textbf{(4)} & \textbf{(5)} & \textbf{(6)} & \textbf{(7)} \\  \hline 
10215-3703 & 0.0320 & LIN & RG & 3.2158 & 10.1102 & - \\
10498-6104 & 0.0507 & LIN & RG & 3.0719 & 10.2670 & - \\
10504-3703 & 0.0230 & LIN & RG & 3.6443 & 10.0380 & - \\
10510-6103 & 0.0195 & LIN & wAGN & 12.2742 & 10.6460 & - \\ 
\end{tabular}
\tablefoot{MaNGA ID (1) is assigned following the convention "[PLATE]-[IFUDESIGN]". Redshift (2) is directly taken from the data reduction processed files. Classifications based on the BPT diagrams (3) and WHAN diagrams (4) are obtained using non-corrected from reddening emission line fluxes. The R\ensuremath{_{50}} petrossian isophote (5) and the stellar (6) and H{\sc I} (7) masses are taken from the NASA-Sloan Atlas (NSA) catalog\footnote{\url{https://www.sdss4.org/dr17/manga/manga-target-selection/nsa/}.}. The complete version of this table is available at the CDS.}
\end{table*}

\begin{table*}
\caption{Optical spectroscopic information for our sample of LINER-like galaxies.}
\label{Sample_lines}
\centering
\begin{tabular}{c c c c c c c} 
\hline\hline
\textbf{MaNGA ID} & \boldmath$W_{H_{\alpha }} [\AA]$ & \boldmath$c \left( H_{\beta} \right) $ & \textbf{[\ion{O}{II}]}\boldmath$\lambda\,\lambda3727,3729$  & ... & \textbf{[\ion{S}{II}]}\boldmath$\lambda6717$ & \textbf{[\ion{S}{II}]}\boldmath$\lambda6731$ \\ \textbf{(1)} & \textbf{(2)} & \textbf{(3)} & \textbf{(4)} & ... & \textbf{(9)} & \textbf{(10)} \\  \hline 
10215-3703 & 2.74 & 0.1197$\pm$0.0871 & 3.146$\pm$0.283 & ... & 1.407$\pm$0.144 & 0.865$\pm$0.100  \\
10498-6104 & 2.5 & 0.4298$\pm$0.1100 & 2.844$\pm$0.357 & ... & 1.028$\pm$0.133 & 0.761$\pm$0.108  \\
10504-3703 & 2.18 & - & 2.858$\pm$0.327 & ... & 0.985$\pm$0.154 & 0.943$\pm$0.153  \\
10510-6103 & 3.87 & 1.4050$\pm$0.1157 & 14.617$\pm$1.595 & ... & 1.458$\pm$0.183 & 1.529$\pm$0.193  \\
\end{tabular}
\tablefoot{Column (1): MaNGA ID. Column (2): equivalent width of H\ensuremath{\alpha}. Column (3): extinction coefficient based on \citet{Howarth_1983} extinction law. Columns (4)-(10): optical emission line ratios [\ion{O}{II}]$\lambda\,\lambda$3727,3729, [\ion{Ne}{iii}]\ensuremath{\lambda }3868\ensuremath{\AA}, [\ion{O}{III}]$\lambda$4959, [\ion{O}{III}]$\lambda$5007, [\ion{N}{II}]$\lambda$6548, [\ion{N}{II}]$\lambda$6584, and [\ion{S}{II}]$\lambda \lambda$6717,6731 corrected from reddening and referred to  H$\beta$. The complete version of this table is available at the CDS.}
\end{table*}

\begin{table*}
\caption{Estimated oxygen abundances in the nuclear region of our sample of LINER-like galaxies based on different grids of photoionization models.}
\label{Sample_oxygen}
\centering
\begin{tabular}{c c c c c c } 
\hline\hline
\textbf{MaNGA ID} & \boldmath$12+log(O/H)_{AGN \ \alpha_{OX} = -0.8}$ & \boldmath$12+log(O/H)_{AGN \ \alpha_{OX} = -1.0}$    & ...   & \boldmath$12+log(O/H)_{pAGB \ T_{eff} = 150kK}$ \\ \textbf{(1)} & \textbf{(2)} & \textbf{(3)}  & ...  & \textbf{(12)} \\  \hline 
10215-3703 & 8.73$\pm$0.12 & 8.56$\pm$0.10  & ... &  8.32$\pm$0.08 \\
10498-6104 & 8.72$\pm$0.24 & 8.66$\pm$0.22  & ...  & 8.24$\pm$0.09 \\
10504-3703 & 8.71$\pm$0.21 & 8.64$\pm$0.20  & ...  & 8.28$\pm$0.08 \\
10510-6103 & 8.73$\pm$0.06 & 8.70$\pm$0.06  & ...  & 8.82$\pm$0.05 \\
\end{tabular}
\tablefoot{Column (1): MaNGA ID. Columns (2)-(12): oxygen abundances for each grid considered. The complete version of this table is available at the CDS.}
\end{table*}

\begin{table*}
\caption{Estimated nitrogen-to-oxygen abundance ratios in the nuclear region of our sample of LINER-like galaxies based on different grids of photoionization models.}
\label{Sample_nitrogen}
\centering
\begin{tabular}{c c c c c c } 
\hline\hline
\textbf{MaNGA ID} & \boldmath$log(N/O)_{AGN \ \alpha_{OX} = -0.8}$ & \boldmath$log(N/O)_{AGN \ \alpha_{OX} = -1.0}$    & ...   & \boldmath$log(N/O)_{pAGB \ T_{eff} = 150kK}$ \\ \textbf{(1)} & \textbf{(2)} & \textbf{(3)}  & ...  & \textbf{(12)} \\  \hline 
10215-3703 & -0.80$\pm$0.15 & -0.74$\pm$0.13 & ... & -0.63$\pm$0.11 \\
10498-6104 & -0.67$\pm$0.11 & -0.67$\pm$0.12 & ... & -0.55$\pm$0.10 \\
10504-3703 & -0.68$\pm$0.14 & -0.64$\pm$0.10 & ... & -0.57$\pm$0.10 \\
10510-6103 & -1.20$\pm$0.13 & -1.14$\pm$0.09 & ... & -1.06$\pm$0.11 \\
\end{tabular}
\tablefoot{Column (1): MaNGA ID. Columns (2)-(12): nitrogen-to-oxygen abundance ratios for each grid considered. The complete version of this table is available at the CDS.}
\end{table*}

\begin{table*}
\caption{Estimated ionization parameters in the nuclear region of our sample of LINER-like galaxies based on different grids of photoionization models.}
\label{Sample_ionization}
\centering
\begin{tabular}{c c c c c c } 
\hline\hline
\textbf{MaNGA ID} & \boldmath$log(U)_{AGN \ \alpha_{OX} = -0.8}$ & \boldmath$log(U)_{AGN \ \alpha_{OX} = -1.0}$    & ...   & \boldmath$log(U)_{pAGB \ T_{eff} = 150kK}$ \\ \textbf{(1)} & \textbf{(2)} & \textbf{(3)}  & ...  & \textbf{(12)} \\  \hline 
10215-3703 & -3.59$\pm$0.08 & -3.71$\pm$0.04 & ... & -3.66$\pm$0.08 \\
10498-6104 & -3.46$\pm$0.08 & -3.53$\pm$0.07 & ... & -3.50$\pm$0.07 \\
10504-3703 & -3.45$\pm$0.11 & -3.53$\pm$0.12 & ... & -3.52$\pm$0.08 \\
10510-6103 & -3.74$\pm$0.05 & -3.78$\pm$0.04 & ... & -3.81$\pm$0.04 \\
\end{tabular}
\tablefoot{Column (1): MaNGA ID. Columns (2)-(12): ionization parameters for each grid considered. The complete version of this table is available at the CDS.}
\end{table*}
\end{document}